\documentclass[10pt,twocolumn,preprintnumbers,nofootinbib,superscriptaddress,aps,prd,showkeys]{revtex4-1}

\usepackage[colorlinks,linkcolor=red,citecolor=blue,urlcolor=blue]{hyperref}
\usepackage{amsmath}  
\usepackage{amssymb}  
\usepackage{theorem}
\usepackage{soul} 
\usepackage{lmodern}
\usepackage{graphicx}
\usepackage{dcolumn}
\usepackage{amssymb,amsmath,bm}
\usepackage{color}
\usepackage{multirow}
\usepackage{enumitem}
\usepackage{mathptmx} 
\usepackage{tensor}
\usepackage{amssymb} 
\usepackage{siunitx} 
\usepackage{lettrine}
\usepackage[normalem]{ulem}
\usepackage{empheq}
\usepackage{comment}


\begin{document}
  \title{Gravitational Wave Birefringence in generalized Palatini Chern Simons}
  
\author{José Perdiguero}
 \email{j.perdiguerogarate@uandresbello.edu}
  \affiliation{Department of Physics and Astronomy, Universidad Andrés Bello, Santiago, Chile}
   \author{Macarena Lagos}
 \email{macarena.lagos.u@unab.cl}
  \affiliation{Department of Physics and Astronomy, Universidad Andrés Bello, Santiago, Chile}

\begin{abstract}
The cosmological propagation of gravitational waves (GWs) can exhibit amplitude and phase polarization distortions when parity symmetry is broken, a phenomenon known as cosmological birefringence. In this paper, we investigate the phenomenology of GW birefringence in a gravitational model $f(R)$ coupled to a dynamical Chern-Simons (dCS) term, analyzed in the metric and  Palatini formalisms. At the background level, this model can lead to dynamical dark energy, while for GW propagation we find that the Palatini formalism predicts both amplitude and velocity birefringence, whereas the metric formalism predicts amplitude birefringence only. We also find that the birefringence effects can either be suppressed or enhanced by the $f(R)$ interactions, depending on the specific form of $f(R)$. Considering three common $f(R)$ models (Hu-Sawicki, Exponential, and Hyperbolic gravity) fit to recent cosmological data, we find that birefringence is enhanced relative to the $f(R) = R$ case, by $1–10\%$ in the Palatini formalism and up to a factor of 2 in the metric formalism. We also translate current GW birefringence constraints to bounds on the dCS coupling within our model. Finally, we show that in our model the birefringence effect grows polynomially with source redshift, in contrast to the linear-distance scaling commonly assumed in phenomenological models of GW birefringence in the current literature.
\end{abstract}


\maketitle

\section{Introduction}
 
The detection of gravitational waves (GWs) from binary black hole (BBH) mergers by the LIGO-Virgo-KAGRA (LVK) collaboration has established a new experimental frontier, opening a landscape for probing the fundamental nature of gravity, and testing theories beyond General Relativity (GR) \cite{LIGOScientific:2016aoc,LIGOScientific:2017zic,LIGOScientific:2018mvr,LIGOScientific:2020ibl,KAGRA:2021vkt}.
Current observations have not yet found significant deviations from the predictions of GR \cite{Johnson-McDaniel:2019zkl,LIGOScientific:2018dkp,LIGOScientific:2019fpa,LIGOScientific:2020tif,LIGOScientific:2021sio,LIGOScientific:2026qni,LIGOScientific:2026sit}, although the associated uncertainties remain large enough that the data cannot yet exclude many modified gravity scenarios. Nonetheless, it is still necessary to consider modified theories of gravity in order to address some of the open problems facing GR. These include the difficulty of quantizing and renormalizing GR \cite{DeWitt:1967yk,DeWitt:1967ub,DeWitt:1967uc,Deser:1974xq} in the high-energy regime, and the need for a fundamental explanation of the dark sector \cite{Zwicky:1937zza,Rubin:1970zza,SupernovaSearchTeam:1998fmf,SupernovaCosmologyProject:1998vns,Sofue:2000jx} (dark matter and dark energy), which together account for approximately $96\%$ of the energy content of the universe today \cite{Planck:2018vyg,DESI:2024mwx}. As a result, GR is understood to be an effective theory of gravity that may require modifications or extensions at very short and very large scales.

Current cosmological observations of the cosmic microwave background (CMB) anisotropies, large-scale structures (LSS), and the expansion history of the universe have provided precise constraints on the parameters of the concordance $\Lambda$CDM model \cite{Planck:2018vyg}. Nonetheless, as precision has increased and more surveys have become available, several tensions have emerged. 
Among these are the measurements of the expansion rate of the universe (known as the Hubble tension) \cite{Di_Valentino_2021}, 
indications of time-varying dark energy from the DESI galaxy survey \cite{DESI:2024mwx}, and possible evidence of cosmological parity violation in the universe \cite{Minami:2020odp,Philcox:2022hkh,Hou:2022wfj}. 

Explaining a possible time-varying dark energy requires modifying or extending the $\Lambda$CDM model. Even though it is still possible within GR to have a dynamical dark energy associated with a minimally coupled scalar field, recent results from DESI \cite{DESI:2024mwx} disfavor this model, since the data indicate that dark energy crosses the phantom divide, which makes a quintessence field unstable \cite{Hu:2004kh}. This opens the door to exploring non-minimal couplings and extensions of GR. There exists a vast landscape of alternative models of gravity \cite{Clifton:2011jh,Ezquiaga:2018btd,Shankaranarayanan:2022wbx}, each motivated by different physical or mathematical considerations. One of the most studied generalizations arises from relaxing the standard Einstein-Hilbert action to an arbitrary function of the Ricci scalar, leading to the class of $f(R)$ theories of gravity \cite{Tsujikawa:2007xu,Cognola:2007zu,Sotiriou:2008rp,DeFelice:2010aj,Olmo:2011uz}. One can go even further and include non-trivial tensor fields, such as the non-metricity tensor and the torsion tensor, both of which vanish in GR, to explore new gravitational dynamics. A map and classification of modified theories of gravity can be found in Refs.~\cite{Cognola:2007zu,Hu:2007nk,Santos:2012vs,Nojiri:2020pqr,Plaza:2025gcv}. Recent studies fitting modified gravity models to CMB and DESI data include \cite{Chudaykin:2024gol,Turyshev:2026ewm,Gonzalez:2026gjd,Mazumdar:2026sdo,Toda:2026yum}.

Regarding parity violation, initial evidence for cosmological parity-violation was found in the CMB \cite{Minami:2020odp}, with further support coming from the four-point correlation function of LSS \cite{Philcox:2022hkh,Hou:2022wfj}. However, these findings remain a subject of debate due to potential systematic effects \cite{Clark:2021kze,Diego-Palazuelos:2022mcp,Philcox:2024mmz}. Forthcoming cosmological surveys such as CMB-S4 \cite{CMB-S4:2016ple}, Rubin \cite{LSSTScience:2009jmu}, Euclid, \cite{Euclid:2021icp} and SKA \cite{Weltman:2018zrl} are expected to provide the precision necessary to determine whether the cosmological parity symmetry is indeed broken.

GWs also offer a powerful and independent avenue for constraining cosmology. For example, the first GW-EM multi-messenger event, GW170817, from a binary neutron star (BNS) merger \cite{LIGOScientific:2017zic} provided the first GW constraints on the expansion rate of the universe \cite{LIGOScientific:2017adf}. GWs can also probe dynamical dark energy through a variety of propagation signatures. These include modifications to the propagation speed of GWs \cite{Wang:2017rpx,Baker:2017hug,Ezquiaga:2017ekz}, amplitude damping such as that predicted in $f(R)$ models \cite{Belgacem:2018lbp, Lagos:2019kds, LISACosmologyWorkingGroup:2019mwx}, and phase distortions accumulated over cosmological distances \cite{Mirshekari:2011yq,Nojiri:2020pqr,Mastrogiovanni:2020gua, Ezquiaga:2021ler}. In parity-violating theories of gravity such as dynamical Chern-Simons (dCS) gravity \cite{Lue:1998mq,Jackiw:2003pm,Alexander:2009tp}, GW propagation exhibits cosmological birefringence, whereby the left- and right-handed circular polarization modes propagate differently, inducing polarization-dependent changes in both the amplitude and phase of the signal. This effect can be constrained using binary black hole (BBH) mergers \cite{Okounkova:2021xjv,Ng:2023jjt,LIGOScientific:2026fcf}, and BNS events with electromagnetic counterparts \cite{Lagos:2024boe}, which provide the sky localization, redshift, and polarization information necessary to disentangle these signatures. Current ground-based detectors from the LVK collaboration, at design sensitivity, are expected to observe GW events up to redshift $z \sim 1$, but next-generation observatories such as Cosmic Explorer (CE) \cite{Evans:2021gyd}, the Einstein Telescope (ET) \cite{ET:2019dnz}, and the space-based detector LISA \cite{LISA:2024hlh} will extend this reach to $z \sim 10$, enabling precision tests of gravity and cosmology across a much broader range of cosmic history.

In this paper, we investigate a model that simultaneously  predicts a time-evolving dark energy and cosmological parity violation in GWs. Specifically, we consider a gravity model that contains a general $f(R)$ function in addition to a dCS coupling. This combination is phenomenologically well motivated: while the $f(R)$ term drives the background expansion and can reproduce dynamical dark energy consistent with DESI observations \cite{Plaza:2025ryz,DOnofrio:2025cuk,Feng:2025cwi,Odintsov:2026fyc}, the dCS term introduces parity violation at the perturbation level, generating GW birefringence testable with current and future detectors. We analyze this modified gravity model using both the metric and Palatini formalisms, which give rise to different phenomenological effects, 
thereby extending the analysis of dCS gravity in \cite{Sulantay:2022sag} valid for $f(R)=R$ to an arbitrary $f(R)$ function.

We calculate the equations of motion of the theory, obtain the cosmological background equations of motion and derive GW propagation equations that lead to amplitude and velocity birefringence. In the late-time universe, the accelerated expansion can be driven by the presence of $f(R)$ instead of a cosmological constant, and in that regime we analyze the phenomenology of GW birefringence. By considering three different $f(R)$ models, we show that GW birefringence can be either enhanced or suppressed relative to the case of $f(R)=R$ \cite{Sulantay:2022sag}. We also discuss how current GW constraints on amplitude birefringence 
\cite{Lagos:2024boe,Ng:2023jjt}
and velocity birefringence \cite{Zhao:2022pun,Wang:2021gqm,Jenks:2023pmk} apply to our model, and finally, given that LVK detectors can observe events up to $z\approx 1$, we illustrate how birefringence effects evolve with redshift $z$ during the dark energy dominated epoch.

The structure of the paper is as follows. In Sec.~\ref{sec:f(R)-CS} we introduce the model, which features a general function $f(R)$ coupled to a dCS term, a scalar field, and matter fields. We derive the field equations using the Palatini formalism. Sec.~\ref{sec:cosmological dynamics} presents the cosmological background equations of motion for the metric, the affine connection and the scalar field. In Sec.~\ref{sec:gravitational waves} we analyze tensor cosmological perturbations, deriving their equations of motion in Fourier space, using the Left/Right circular polarization basis. We apply consistency limits to verify that our formulation recovers the standard results from both GR and GR with dCS corrections. In Sec.~\ref{sec:phenomenology}, we present the friction $\Xi_{L,R}$ and the angular velocity $\omega_{L,R}^2$ of the GW in a de-Sitter type of universe; we then introduce toy models of $f(R)$ and study whether the resulting contribution enhances or suppresses the birefringence effect, comparing the Palatini and metric formalisms. We then derive constraints according to recent cosmological data, and finally, present the evolution of $\Xi_{L,R}$ and $\omega_{L,R}^2$ as a function of the redshift, which requires solving the background equations. Our conclusions and final discussion are presented in Sec.~\ref{sec:discussion}. For completeness, we include three appendices. Appendix~\ref{sec:appendix A, solution} presents the explicit solution to the perturbed affine connection equation, Appendix~\ref{sec:appendix B, series expansion} presents the coefficient solutions for the gravitational wave equation, and Appendix~\ref{sec:appendix C, metric formalism} briefly illustrates the metric formalism.

Throughout this paper we set the speed of light to unity, $c=1$, and parentheses denote symmetrization, $A_{(\mu}B_{\nu)} = (A_\mu B_\nu + A_\nu B_\mu)/2$, whereas square bracket denote anti-symmetrization, $A_{[\mu}B_{\nu]} = (A_\mu B_\nu - A_\nu B_\mu)/2$.

\section{The gravity model}
\label{sec:f(R)-CS}

A parity transformation corresponds to a point reflection of the spatial coordinates, and a theory is said to be parity invariant if its physical laws remain unchanged under such a point reflection. This is the case of GR. Since GR is parity invariant, breaking this symmetry requires the introduction of new physics.

In particular, there are several approaches proposed in the literature to describe a parity-violating theories of gravity. Examples include models with higher-derivative scalar field interactions, such as ghost-free scalar-tensor theories \cite{Crisostomi:2017ugk,Nishizawa:2018srh,Zhao:2019xmm,Qiao:2021fwi}; the symmetric teleparallel equivalent of GR \cite{Conroy:2019ibo}, which is a non-metric theory formulated in terms of the non-metricity tensor; and Ho\v{r}ava-Lifshitz gravity \cite{Zhu:2013fja}, which breaks Lorentz invariance at higher energies in order to be a power-counting renormalizable. However, if one requires the theory to be at most quadratic in curvature (motivated by particle physics, string theory and loop quantum gravity \cite{Alexander:2009tp}) and coupled to a scalar field (which need not appear linearly), the simplest parity-violating term that can be introduced is the dynamical Chern-Simons (dCS) term. This term is defined by the contraction of the Riemann curvature tensor with its dual (where the dual operation is responsible for breaking parity symmetry) coupled to a scalar field \cite{Alexander:2009tp,Lue:1998mq,Jackiw:2003pm}. Extensions to dCS gravity have been developed within the Palatini formalism, known as Palatini dCS gravity \cite{Sulantay:2022sag}. Extensions having a connection with a non-vanishing torsion field are referred to as torsional Chern-Simons \cite{Boudet:2022nub,Bombacigno:2022naf}, and there are also models that preserve projective symmetry, which require the inclusion of additional terms in the dCS action \cite{Boudet:2022wmb}. 

Using the metric formalism, the equations of motion of dCS gravity yield third-order derivatives of the metric tensor. This can lead to an ill-posed problem, an issue explored in \cite{Delsate:2014hba}. Several approaches have been developed to address this problem. The first is to adopt a first-order formalism, where the spacetime metric is defined in terms of a vielbein and a spin connection, which yields first-order derivatives in the equations of motion \cite{BottaCantcheff:2008pii}. A second solution utilizes the Palatini formalism, where the metric tensor and the affine connection are treated as independent fields on the manifold, with their relationship  established dynamically through the equations of motion. This results in equations of motion with at most first-order derivatives in the metric tensor and second-order derivatives in the affine connection \cite{Ferraris:1982wci,Olmo:2011uz}. 

The dCS model is one of the most well-known parity-violating extensions of GR. In this theory, GWs are predicted to exhibit parity-violating effects, namely amplitude and velocity birefringence. Gravitational waves in GR have two linear polarization modes, commonly known as the ``plus" $h_{+}$ and ``cross'' $h_{\times}$ \cite{Misner:1973prb,Flanagan:2005yc,Maggiore:2007ulw}, which describe how spacetime oscillates as a GW passes. In the $\Lambda$CDM model, the two polarization modes obey the same propagation equation; therefore, the relative amplitude and phase of these modes do not change from the source to the observer. However, in dCS, and in any parity-violating theory of gravity, the linear polarization modes are coupled to each other. These linear modes can be decoupled by performing an appropriate transformation $h_{+,\times} \to h_{L,R}$, where the new modes are known as the circular polarization basis, with polarizations ``left-handed" $h_{L}$ and ``right-handed" $h_{R}$ \cite{Jackiw:2003pm,Alexander:2009tp,Odintsov:2022hxu,Sulantay:2022sag, Jenks:2023pmk}. This transformation leads to a different sign in the propagation equations of motion for the circular polarization modes, giving rise to the birefringence phenomenon. In this case, the circular polarizations evolve differently during cosmological propagation and, therefore, their relative amplitude and phase change when traveling from the source to the observer.

A natural extension of the dCS model can be achieved by generalizing the Einstein-Hilbert action to an arbitrary function of the scalar curvature, leading to the class of theories known as $f(R)$ gravity. This approach has been explored in the literature within the metric formalism \cite{Odintsov:2022hxu}, and applied to the study of primordial GWs for the specific case $f(R) \sim R + R^2$.

Extending dCS gravity to an $f(R)$ framework is phenomenologically well motivated in light of recent cosmological tensions, which increasingly suggest that dark energy is not a constant but instead evolves with time  \cite{Cortes:2024lgw,DESI:2025fii,DESI:2025zgx}.  
While quintessence scalar fields provide a simple mechanism to account for the accelerated expansion of the universe with a time-evolving energy density, they face difficulties in consistently fitting DESI observations \cite{Chakraborty:2025syu,Wolf:2025jed}. In this context, the Palatini formulation of dCS gravity is effectively equivalent, at the background cosmological level, to introducing a single quintessence scalar  degree of freedom relative to $\Lambda$CDM, thereby limiting its ability to reproduce the dynamical behavior of the dark energy sector. By contrast, extending the theory to an $f(R)$ framework naturally incorporates additional geometric degrees of freedom, enabling a richer phenomenology that can emulate dynamical dark energy in the expansion history of the universe \cite{Plaza:2025gcv}.

In this work, we primarily employ the Palatini formalism, in which the resulting field equations remain of second order, ensuring a well-posed formulation of the theory. We follow and extend the results of \cite{Sulantay:2022sag} to $f(R)$ gravity. As shown therein, the Palatini formalism gives rise to both amplitude and velocity birefringence, in contrast to the metric formalism, which produces only amplitude birefringence \cite{Sulantay:2022sag}. For completeness, we also briefly present the metric formalism in Appendix~\ref{sec:appendix C, metric formalism}, since its derivation is straightforward and has been considered in \cite{Odintsov:2022hxu}.

\subsection{The action}

Let us begin by writing the action of a gravity theory containing an arbitrary function of the scalar curvature, denoted by $f(R)$, with the addition of a parity-violating dCS term coupled to a scalar field $\vartheta$, as well as the matter fields. The total action is given by
\begin{align}
\label{eq: action}
    S&[g,\Gamma, \vartheta,\psi]  = \kappa \int \mathrm{d}^4 x \sqrt{-g} f(R)  +  \frac {\alpha}{4} \int \mathrm{d}^4 x \vartheta \phantom{}^* R R\nonumber\\
    & + \frac{\beta}{2}\int \mathrm{d}^4x \sqrt{-g}\left(\nabla^\mu\vartheta\nabla_\mu\vartheta -2V(\vartheta)\right) + S_{mat}[g,\psi], 
\end{align}
where $\kappa$ is a constant with suitable dimensions, such that when $f(R) = R$, it takes the expected value of $\kappa^{-1} = 16\pi G$. The coupling constants $\alpha$ and $\beta$ are real constant parameters: the former controls the strength of the dCS contribution, while the latter determines the coupling of the scalar field.

Note that the action in Eq.~\eqref{eq: action} is written in the Palatini formalism, meaning it depends on the metric tensor and affine connection separately, in contrast to the metric formalism, where the gravitational action would only depend on the metric tensor.

The first term in Eq.~\eqref{eq: action} is a generalization of the Einstein-Hilbert action, where the gravitational sector is defined by an arbitrary function of the scalar curvature $f(R)$. This action extends the one considered in \cite{Sulantay:2022sag} through the inclusion of the $f(R)$ term.

The second term in Eq.~\eqref{eq: action} corresponds to the CS contribution coupled to a scalar field; this configuration introduces dynamics to the CS and is known as dCS. We refer the reader to
\cite{Yunes:2009hc,Alexander:2009tp,Furtado:2010ys,Peng:2022ttg,Raushan:2023pdv,Alexander:2024vav} for some of its applications. The CS term is constructed by the contraction of the Riemann curvature tensor with its dual 
\begin{equation}
\label{CS}
    \phantom{}^* R R  = \frac{1}{2}\epsilon^{\mu\nu\rho\sigma} R^{\alpha}{}_{\beta\rho\sigma}R^{\beta}{}_{\alpha\mu\nu},
\end{equation}
where $\epsilon^{\mu\nu\rho\sigma}$ is the totally antisymmetric Levi-Civita tensor density responsible for the parity violation of the model. To see this explicitly, consider its transformation under an arbitrary coordinate change
\begin{equation}
    \tilde{\epsilon}^{\alpha\beta\gamma\delta} = 
    J \frac{\partial x^\mu}{\partial x'^\alpha}
    \frac{\partial x^\nu}{\partial x'^\beta}
    \frac{\partial x^\rho}{\partial x'^\gamma}
    \frac{\partial x^\sigma}{\partial x'^\delta}
    \epsilon^{\mu\nu\rho\sigma},
\end{equation}
where $J$ is the determinant of the Jacobian transformation matrix. Under a parity transformation $\hat{P}:x \to -x$, one finds $J = -1$. Each partial derivative contributes a factor of $-1$, however, since the expression contains four derivatives, their combined contribution is $(-1)^4 = +1$. The overall minus sign comes from the determinant of the Jacobian matrix, causing the tensor density to change sign. This property leads to a parity violation in the propagation of the GWs (tensor perturbations), although the action remains parity invariant.

In the context of cosmology, the dCS term does not modify the equations of motion of the FRW background, due to its symmetries. However, it introduces non-trivial effects at the perturbation level, particularly in the tensor perturbations associated with GWs. As a consequence, the amplitude and phase of the GW propagation are polarization-dependent, a phenomenon known as birefringence.

Finally, we note that the CS term can also be written as a topological current 
\begin{equation}
    K^{\mu} = 2\epsilon^{\mu\alpha\beta\gamma}\left(\frac{1}{2}\Gamma^{\sigma}{}_{\alpha\tau}\partial_{\beta}\Gamma^{\tau}{}_{\gamma\sigma} + \frac{1}{3}\Gamma^{\sigma}{}_{\alpha\tau}\Gamma^{\tau}{}_{\beta\eta}\Gamma^{\eta}{}_{\gamma\sigma}\right),
\end{equation}
satisfying the relation
\begin{equation}
    \partial_{\mu}K^{\mu} \equiv \frac{1}{2}\phantom{}^*RR,
\end{equation}
both of which are equivalent representations of the CS term. The divergence of the topological current is the gravitational Pontryagin density. A proof of the above result can be found in \cite{Alexander:2009tp}.

The third term in Eq.~\eqref{eq: action} is the action of a scalar field minimally coupled to gravity, with an arbitrary potential $V(\vartheta)$. Unlike the dCS contribution, the scalar field action does not break parity symmetry; yet its field equations do due to the presence of the dCS term. 

The final term in Eq.~\eqref{eq: action} represents the matter action (with a representative matter field $\psi$), where matter is assumed to be minimally coupled to gravity, as in GR. As a result, this term depends on the metric tensor and not the connection. 

\subsection{Field equations}\label{sec:general_eqns}

In the Palatini formalism of gravity, the metric tensor and the affine connection are treated as independent fields, whose relationship is determined by the equations of motion. Depending on the structure of the $f(R)$ theory, this relationship may yield the Levi-Civita connection or a more general connection. In this framework, the Riemann curvature tensor is defined through the commutator of covariant derivatives as follows
\begin{equation}
    R^{\alpha}{}_{\beta \mu \nu}(\Gamma) = \partial_{\mu}\Gamma^{\alpha}{}_{\nu\beta}-\partial_{\nu}\Gamma^{\alpha}{}_{\mu\beta}+\Gamma^{\alpha}{}_{\mu\lambda}\Gamma^{\lambda}{}_{\nu\beta}-\Gamma^{\alpha}{}_{\nu\lambda}\Gamma^{\lambda}{}_{\mu\beta}.
\end{equation}
Note that the Riemann tensor naturally carries an upper index. Contracting its first and third indices leads to the Ricci tensor, as usual,
\begin{equation}
    R_{\mu\nu}(\Gamma) = R^{\alpha}{}_{\mu\alpha\nu},
\end{equation}
and its contraction with the inverse metric tensor gives the scalar curvature
\begin{equation}
    R(g, \Gamma) = g^{\mu\nu}{R}_{\mu\nu}.
\end{equation}
Unlike the Riemann curvature tensor and the Ricci tensor, the scalar curvature requires the existence of a metric tensor on the manifold.

The field equations are obtained by applying the variational principle to the action in Eq.~\eqref{eq: action}. A variation with respect to the inverse metric tensor $ g^{\mu\nu}$ yields
\begin{equation}
\label{feq metric}
    f_R\,R_{\mu\nu} - \frac{1}{2}g_{\mu\nu}f = \kappa T_{\mu\nu},
\end{equation}
where $f_R = \frac{\partial f(R)}{\partial R}$, which generalizes Einstein's field equations in the Palatini formalism. Setting $f(R) = R$ allows us to recover the standard results from GR. 

In addition, $T_{\mu\nu}$ is the total stress-energy tensor, given by the sum of two contributions: one coming from the dynamical scalar field $\vartheta$, and another from the standard matter sector
\begin{equation}
    T_{\mu \nu} = T^{(\vartheta)}_{\mu \nu} + T^{(\text{mat})}_{\mu \nu},
\end{equation}
where each $T^{(i)}_{\mu\nu}$ with $i = \vartheta, \text{mat}$ is defined as
\begin{align}
    T^{(\text{mat})}_{\mu \nu} &\equiv  -\frac{2}{\sqrt{-g}} \frac{\delta \mathcal{L}_{mat}}{\delta g^{\mu \nu}}, &  
    T^{(\vartheta)}_{\mu \nu} \equiv-\frac{2}{\sqrt{-g}}  \frac{\delta \mathcal{L}_{\vartheta}}{\delta g^{\mu \nu}}.
\end{align}
The contribution from the scalar field takes the standard form for a minimally coupled field
\begin{equation}
\label{eq: scalar field tensor}
    T^{(\vartheta)}_{\mu \nu} = \beta\left(\nabla_{\mu}\vartheta\nabla_{\nu}\vartheta-g_{\mu \nu} \left(\frac{1}{2}\nabla_{\lambda}\vartheta \nabla^{\lambda}\vartheta + V(\vartheta)\right)\right),
\end{equation}
while the matter contribution will be taken to be that of a perfect fluid
\begin{equation}
    T^{(\text{mat})}_{\mu\nu} = (\rho + p)u_\mu u_\nu + p g_{\mu\nu}.
\end{equation}
Here, $\rho$ is the energy density of the fluid, $p$ its  pressure, and $u_\mu$ its 4-velocity vector satisfying the constraint $g^{\mu\nu}u_{\mu}u_{\nu} = u^{\mu}u_{\mu} = -1$. For cosmological applications, we consider an equation of state $p = w \rho$, where the parameter $w$ specifies the type of matter: radiation ($w=1/3$) or non-relativistic matter ($w=0$). One could also include a cosmological constant ($w=-1$), but in this paper the function $f(R)$ will be responsible for the late-time accelerated expansion of the universe and hence we will not add an explicit cosmological constant term. 

Next, varying the action with respect to the affine connection $\Gamma^\lambda{}_{\mu\nu}$ leads to 
\begin{align}
\label{feq connection}
     & \nabla_{\lambda}\left(\sqrt{-g}g^{\mu\nu}f_R\right) -\delta^{(\mu}_{\lambda}\nabla_{\rho}\left(\sqrt{-g}g^{\nu)\rho}f_R\right) \nonumber \\
     & = \alpha \frac{\epsilon^{\delta\gamma\rho(\mu}}{\sqrt{-g}}R^{\nu)}{}_{\lambda\delta\gamma}\nabla_{\rho}\vartheta, 
\end{align}
which determines the relationship between the affine connection and the metric tensor, subject to a dCS correction. Note that, for a vanishing dCS coupling ($\alpha \to 0$) and with $f(R) = R$, the Levi-Civita connection is recovered, as one would expect. For the purposes of this work, we assume a torsion-free connection; therefore, the affine connection is symmetric in its two lower indices.

A variation with respect to the scalar field $\vartheta$ gives
\begin{equation}
\label{feq scalar}
  \beta \Box_g \vartheta -\beta \frac{\mathrm{d} V(\vartheta)}{\mathrm{d} \vartheta}=-  \frac{\alpha}{4}\:^{*}R R ,
\end{equation}
where $\Box_g$ denotes the d'Alembertian operator defined as 
\begin{equation}
    \Box_g = \frac{1}{\sqrt{-g}}\partial_\alpha\left(\sqrt{-g}g^{\alpha\beta}\partial_\beta\right).
\end{equation}

In contrast to the purely metric formalism, where only the metric tensor and the scalar field yield equations of motion, in the Palatini formalism an additional equation, Eq.~\eqref{feq connection}, associated with the independent affine connection is introduced.

The equation associated with the matter fields can be written as a conservation law
\begin{equation}
    \nabla^\mu T^{(\text{mat})}_{\mu \nu} = 0.
\end{equation}
This result can be derived from the equations of motion; for a detailed review see \cite{Koivisto:2005yk}.

As a comparison, in the metric formalism the manifold is endowed with only a metric structure $\mathcal{M}(g)$, where $g_{\mu\nu}$ allows us to define the notion of distance and the connection, which in turn defines the notion of parallelism via the Levi-Civita connection $\Gamma^{\mu}_{\alpha\beta}$, whose coefficients are given by the Christoffel symbols
\begin{equation}
    \Gamma^{\mu}{}_{\alpha\beta} = \frac{1}{2}g^{\mu\rho}\left(\partial_\alpha g_{\beta\rho}  + \partial_\beta g_{\alpha\rho} - \partial_\rho g_{\alpha\beta}\right).
\end{equation}
To find the field equations, it is necessary to vary the action $S = S[g,\vartheta,\psi]$ with respect to the metric tensor. In this formalism, the scalar field and matter equations are the same as the ones obtained in the Palatini formalism, but the metric equation is different. The resulting equations of motion are presented in Appendix~\ref{sec:appendix C, metric formalism}.

Finally, comparing both the metric and Palatini formalisms, the former yields higher-order derivatives of the metric tensor, whereas the latter yields only first-order derivatives of the metric tensor and the affine connection. Therefore, the metric formalism will lead to an ill-posed problem \cite{Delsate:2014hba}, which is avoided when using the Palatini formalism \cite{Ferraris:1982wci}. 

\section{Cosmological Background}
\label{sec:cosmological dynamics}

In this section, we present the cosmological ansatz for the fundamental fields of our model, namely, the metric tensor $g_{\mu\nu}$, the affine connection $\Gamma^{\mu}{}_{\rho\sigma}$, the scalar field $\vartheta$, and the matter fluid, together with their corresponding field equations.

\subsection{Ansatz}

An ansatz for an arbitrary object can be constructed by requiring that its Lie derivative vanish along the Killing vectors that generate the desired symmetries. In our case, these correspond to isotropy (rotations) and homogeneity (translations). Imposing these conditions on the symmetric part of the affine connection yields
\begin{align}
    \mathcal{L}_{\xi} \Gamma^{\alpha}{}_{\mu\nu} & = 0.
\end{align}
The above conditions lead to a system of differential equations that must be solved simultaneously. The symmetric part of $\Gamma^{\alpha}{}_{\mu\nu}$ is characterized by three independent functions
\begin{align}
\label{cosmo connection}
    \Gamma^{t}{}_{tt}& = j(\eta), & 
    \Gamma^{t}{}_{ij}& = l(\eta) \delta_{ij}, & 
    \Gamma^{i}{}_{tj}& = b(\eta)\delta^i_j,
\end{align}
A detailed computation of the connection coefficients can be found in Refs.~\cite{Castillo-Felisola:2018jfp,Castillo-Felisola:2019thp,Castillo-Felisola:2024xil}. Note that, while the methodology used here differs from that of \cite{Sulantay:2022sag}, we obtain the same connection ansatz. Although this is a systematic method to find an ansatz for a tensor of rank $(p,q)$, the remaining fields---such as the metric tensor, scalar field, and the matter field--- have well-known structures in the literature, and therefore do not require this algorithm.

The metric tensor $g_{\mu\nu}$ compatible with the symmetries of the cosmological principle contains a single conformally time-dependent function $a = a(\eta)$, known as the scale factor. For a spatially-flat universe, the line element reads
\begin{equation}
\label{conformal metric}
   \mathrm{d}s^2 = a^{2}(\eta)\left(-\mathrm{d}\eta^2 + \delta_{ij}\mathrm{d}x^{i}\mathrm{d}x^{j}\right).
\end{equation}

The scalar field $\vartheta$ depends only on time
\begin{equation}
\label{cosmo scalar}
    \vartheta = \vartheta(\eta).
\end{equation}

Finally, the matter contribution $T^{mat}_{\mu\nu}$ is modeled as a perfect fluid, which in the cosmological framework takes the form 
\begin{equation}
\label{FRWfluid}
    T^{mat}_{\mu \nu} =\left( \rho(\eta) + p(\eta)  \right) u_{\mu}u_{\nu} + p(\eta) g_{\mu \nu}, 
\end{equation}
where $u_{\mu}=(a,0,0,0)$ is the $4$-velocity of the fluid, assumed to be at rest.

In the metric formalism, the independent fields are the metric tensor, the scalar field, and the matter field, whose forms are the same as those defined in the Palatini formalism, see Eqs.~\eqref{conformal metric},~\eqref{cosmo scalar}, and ~\eqref{FRWfluid} respectively. In this formalism, however, the connection is taken to be the Levi-Civita connection, which is fully determined by the metric tensor and therefore does not require an ansatz.

\subsection{Equations of Motion}

To begin our analysis, let us focus on the affine connection equation, Eq.~\eqref{feq connection}. Under the cosmological ansatz, the term coming from the dCS contribution vanishes, leading to the following three equations
\begin{align}
    3abf_R - ajf_R - 2f_Ra' - af'_R & = 0,\\
    abf_R + ajf_R - 2f_Ra' - af'_R & = 0,\\
    b - l & = 0,
\end{align}
where $'$ denotes derivatives with respect to conformal time. The system admits an analytical solution, which is given by
\begin{align}
\label{j connection}
    j(\eta) & = \mathcal{H} + \frac{1}{2}\frac{f'_R}{f_R}, &
    l(\eta) & = b(\eta) = j(\eta),
\end{align}
where $ \mathcal{H}=a'/a$ is the conformal Hubble factor.
The presence of an arbitrary function $f(R)$ at the action level translates into a modification of the affine connection with respect to the Levi-Civita one, with the connection coefficients acquiring an additional contribution proportional to $f_R$ and its derivative. As a consistency check, by fixing $f(R) = R$ we recover the standard Levi-Civita connection.

With the solution to the connection equation in hand, it is straightforward to compute the metric equation, Eq.~\eqref{feq metric}, yielding two differential equations
\begin{align}
    \kappa a^2\rho - \frac{1}{2}a^2f - \frac{3f_Ra'^2}{a^2} + \frac{3f_Ra''}{a} - \frac{3(f'_R)^2}{2f_R} + \frac{3}{2}f''_R & = 0,\\
    -\kappa a^2 p - \frac{1}{2}a^2f + \frac{f_Ra'^2}{a^2} + \frac{f_Ra''}{a} + \frac{2 a'f'_R}{a} + \frac{1}{2}f''_R & = 0.
\end{align}
These two equations can be combined by solving the first one for $a''(\eta)$ and substituting the result into the second one, leading to
\begin{equation}
\label{frw equation}
    \left(\mathcal{H} + \frac{1}{2}\frac{f'_R}{f_R}\right)^2 = \frac{a^2}{6}\frac{\kappa \left(\rho + 3p\right)}{f_R} + \frac{a^2}{6} \frac{f}{f_R}.
\end{equation}
This is the modified Friedmann equation for a spatially flat universe, matching the results from previous works in $f(R)$ theories \cite{Kremer:2004bf,Amarzguioui:2005zq,Sotiriou:2008rp}. We emphasize that $f_R = \frac{\partial f(R)}{\partial R}$ whereas $f'_{R} = \frac{\partial f_R(R)}{\partial \eta}$.

The scalar field equation, computed using only the inverse metric tensor and partial derivatives, leads to the standard differential equation
\begin{equation}
\label{scalar field dynamics}
    \vartheta'' + 2\mathcal{H}\vartheta' + a^2\partial_\vartheta V(\vartheta) = 0,
\end{equation}
with the CS contribution vanishing, leaving the classical Klein-Gordon equation where $\partial_\vartheta$ denotes a derivative with respect to $\vartheta$.

Although Eq.~\eqref{j connection} solves the affine connection equations, yielding a non Levi-Civita connection, it is possible to reformulate the metric equation as if one were working with a Levi-Civita connection. In the cosmological background, the right-hand side of Eq.~\eqref{feq connection} vanishes identically, leading to a simplified equation of motion for the connection
\begin{equation}
\label{eq: connection simplified}
     \nabla_{\lambda}\left(\sqrt{-g}g^{\mu\nu}f_R(R)\right) -\delta^{(\mu}_{\lambda}\nabla_{\rho}\left(\sqrt{-g}g^{\nu)\rho}f_R(R)\right) = 0.
\end{equation}
Taking the trace of the above equation yields
\begin{equation}
    \nabla_{\mu}\left(\sqrt{-g}g^{\mu\nu}f_R(R)\right) = 0,
\end{equation}
from which one concludes that the second term in Eq.~\eqref{eq: connection simplified} vanishes, giving
\begin{equation}
    \nabla_{\alpha}\left(\sqrt{-g}g^{\mu\nu}f_R(R)\right) = 0.
\end{equation}
The solution of the above equation is well known. By introducing a conformal metric $\tilde{g}_{\mu\nu}$ such that
\begin{equation}
\label{conformal metric h}
    \tilde{g}_{\mu\nu} = \Omega^{2}g_{\mu\nu},
\end{equation}
where $\Omega^{2} = f_R(R)$, we can rewrite the previous equation as 
\begin{equation}
    \nabla_{\alpha}\left(\sqrt{-\tilde{g}}\,\tilde{g}^{\mu\nu}\right) = 0.
\end{equation}
This condition implies that the connection must be the Levi-Civita connection associated with the conformal metric $\tilde{g}_{\mu\nu}$
\begin{equation}
\label{conformal connection}
    \Gamma^{\mu}{}_{\alpha\beta}(\tilde{g}) = \frac{1}{2}\tilde{g}^{\mu\rho}\left(\partial_{\alpha}\tilde{g}_{\beta\rho} + \partial_{\beta}\tilde{g}_{\alpha\rho} - \partial_{\rho}\tilde{g}_{\alpha\beta}\right).
\end{equation}
Substituting Eq.~\eqref{conformal metric h} into \eqref{conformal connection}, one can express the Levi-Civita connection $\Gamma(\tilde{g})$ in terms of the metric $g_{\mu\nu}$ as
\begin{align}
    \Gamma^{\mu}{}_{\alpha\beta}(\tilde{g}) & = \Gamma^{\mu}{}_{\alpha\beta}(g) + \frac{1}{2f_R}\bigg(\delta^{\mu}_{\alpha}\partial_{\beta}f_R 
    + \delta^{\mu}_{\beta}\partial_{\alpha}f_R  - g_{\alpha\beta}\partial^{\mu}f_R\bigg),
\end{align}
where $\Gamma^{\mu}{}_{\alpha\beta}(g)$ denotes the Christoffel symbols associated with the metric $g_{\mu\nu}$. Similarly, the Ricci tensor and the scalar curvature can be written solely in terms of $g_{\mu\nu}$ as
\begin{align}
\label{ricci conformal}
    R_{\mu\nu}(\tilde{g}) & = R_{\mu\nu}(g) + \frac{3}{2f^2_R}\left(\nabla_\mu f_R\nabla_{\nu}f_R\right) \nonumber\\
    & - \frac{1}{f_R}\left(\nabla_\mu\nabla_\nu f_R - \frac{1}{2}g_{\mu\nu}\square f_R\right), \\
\label{scalar conformal}
    R(\tilde{g}) & = R(g) + \frac{3}{2f_R^2}\nabla_\mu f_R\nabla^{\mu}f_R - \frac{3}{f_R}\square f_R,
\end{align}
where $R_{\mu\nu}(g)$ and $R(g)$ are computed from the metric tensor $g_{\mu\nu}$ and its corresponding Christoffel symbols $\Gamma^{\mu}{}_{\alpha\beta}(g)$. Replacing Eqs.~\eqref{ricci conformal} and \eqref{scalar conformal} into Eq.~\eqref{feq metric}, the metric field equation can be expressed entirely in terms of the original metric tensor $g_{\mu\nu}$ and the energy-momentum tensor
\begin{align}
\label{feq metric emt}
    R_{\mu\nu}(g) - \frac{1}{2}g_{\mu\nu}R(g) = \kappa\frac{T_{\mu\nu}}{f_R} - g_{\mu\nu}\frac{Rf_R - f}{2f_R} \nonumber\\
    -\frac{3}{2f_R^2}\left(\partial_{\mu}f_R\partial_{\nu}f_R - \frac{1}{2}g_{\mu\nu}(\partial f_R)^2\right) \nonumber \\
    + \frac{1}{f_R}\left(\nabla_{\mu}\nabla_{\nu}f_R - g_{\mu\nu}\square f_R\right),
\end{align}
where $R_{\mu\nu}(g)$, $R(g)$ and $\nabla_{\mu}\nabla_{\nu}f_R$ are computed using the Christoffel symbols associated with the metric $g_{\mu\nu}$. The scalar curvature $R$ and $f_R$ must be understood as functions of the trace of the energy-momentum tensor, which can be obtained by taking the trace of Eq.~\eqref{feq metric}
\begin{equation}
\label{eq: trace palatini}
    f_R R - 2f = \kappa T.
\end{equation}
Therefore, Eq.~\eqref{feq metric emt} is written entirely in terms of the metric tensor $g_{\mu\nu}$ and the energy-momentum tensor, without reference to the independent affine connection. A straightforward computation using Eq.~\eqref{feq metric emt} yields the same equation as Eq.~\eqref{frw equation}. Details of this calculation and a more general discussion on this can be found in  \cite{Sotiriou:2008rp,Olmo:2011uz,Meng:2004wg,Kremer:2004bf,Amarzguioui:2005zq,Sotiriou:2006qn}

For later analyses, it will be useful to introduce an alternative version of the Friedmann modified equation (\ref{frw equation}), which includes contributions coming from matter fields, the arbitrary function $f(R)$, and the scalar field $\vartheta$. Taking the trace of the metric equation yields
\begin{equation}
\label{trace eq derivative}
    R' = \frac{\kappa T'}{Rf_{RR} - f_R},
\end{equation}
where $T$ is the trace of the energy-momentum tensor, which now has contributions from non-relativistic matter, radiation matter and the scalar field. The trace is written as 
\begin{equation}
    T = -\rho_m + 2\left(\frac{\vartheta'^2}{a^2} - V\right) = -\rho_m + T_\vartheta.
\end{equation}
Taking the conformal time derivatives and using the continuity equation for the non-relativistic matter field gives
\begin{equation}
    T' = 3\mathcal{H}\rho_m + T'_\vartheta.
\end{equation}
Substituting the above result into Eq.~\eqref{trace eq derivative} yields
\begin{equation}
    R' = \frac{\kappa}{Rf_{RR} - f_R}\left( 3\mathcal{H}\rho_m + T'_\vartheta\right).
\end{equation}
Starting from Eq.~\eqref{frw equation}, one then finds
\begin{equation}
    \left(\frac{\mathcal{H}}{H_0}\right)^2 = \frac{a^2}{6f_R \xi^2}\left(3\Omega_{m0}(1+z)^3 + 6\Omega_{r0}(1 +z)^4 + 3\Omega_{\vartheta 0} + \frac{f(R)}{H_0^2}\right),
\end{equation}
where $\Omega_{\vartheta0} = \frac{2\kappa\beta}{H_0^2}\left(\frac{\vartheta'^2}{a^2} - 2 V\right)$ and the variable $\xi$ is defined as 
\begin{equation}
    \xi = 1 + \frac{1}{2} \frac{f_{RR}}{f_{R}(R f_{RR} - f_R)}\left(9H_0^2\Omega_{m0}(1 + z)^3 + \frac{\kappa T'_\vartheta}{\mathcal{H}}\right),
\end{equation}
which generalizes the previous results by including the effects of the scalar field $\vartheta$ and its potential. We highlight that no cosmological constant is introduced in this formulation of gravity; instead, the role of dark energy is assumed to be effectively played by $f(R)$. The scalar field is minimally coupled and could also be considered to contribute as dark energy; however, according to recent results \cite{Wolf:2024eph,Wolf:2025jed}, dark energy evolves with time, crossing the phantom divide which suggests it is not well described by a minimally coupled scalar field. In Sec. \ref{sec:phenomenology}, we will thus consider the case where $f(R)$ drives the accelerated expansion of the universe, and the scalar field makes a negligible contribution at late times, that is, $\Omega_{\theta}(z=0)\ll 1$.

Finally, in contrast to the Palatini formalism, there is no independent equation for the connection, since it is fully determined by the metric tensor via the Christoffel symbols. Consequently, there are only two equations coming from the metric tensor, and one equation from the scalar field. The latter is the same as that obtained in the Palatini formalism, as expected, since the scalar field action takes the same form in both formalisms. The explicit expressions for the metric formalism background cosmological equations are in Appendix~\ref{sec:appendix C, metric formalism}.

\section{Gravitational Waves}
\label{sec:gravitational waves}

GWs are introduced into the field equations through tensor perturbations of the metric tensor and the affine connection. There are two main consequences of introducing such perturbations: first, the breaking of the symmetries implied by the cosmological principle (rotations and translations); and second, the appearance of non-trivial contributions arising from the CS term, leading to amplitude and velocity birefringence. 

The metric tensor is perturbed as follows
\begin{equation}
    g_{\mu \nu}= \bar{g}_{\mu \nu}(\eta) + \delta g_{\mu \nu}(\eta, \vec{x}),
\end{equation}
where $\bar{g}_{\mu \nu}(\eta)$ is the background metric, defined in Eq.~\eqref{conformal metric}, and  $\delta g_{\mu \nu}(\eta, \vec{x})$ is a small perturbation  such that higher-order terms can be neglected.

Since we are interested in GWs, we perform a Scalar-Vector-Tensor (SVT) decomposition of $\delta g_{\mu\nu}$ and retain only the tensor perturbations. For reviews on the SVT decomposition, see \cite{Kodama:1984ziu,Mukhanov:1990me,Liddle:2000cg,Malik:2008im}. The new line element, including both the background metric and the tensor perturbations, is given by 
\begin{equation}
\label{line element}
    \mathrm{d}s^2 = a^2(\eta)\left[-\mathrm{d}\eta^2 +\left(\delta_{ij} + h_{ij}\mathrm{d}x^{i}\mathrm{d}x^{j}\right)\right],
\end{equation}
where $h_{ij}$ represents the tensor perturbation, which must satisfy the following transverse and traceless conditions
\begin{align}
    \partial^{i}h_{ij} & = 0 & h^{j}{}_{j} & = 0,
\end{align}
where indices are raised with a flat $\delta_{ij}$ spatial metric.
The tensor perturbation $h_{ij}$ contains only two physical degrees of freedom, corresponding to the two polarization modes of GWs, commonly denoted as plus $+$ and cross $\times$, in a linear basis. This follows from $h_{ij}$ being symmetric (six independent components), the transverse condition $\partial^i h_{ij} = 0$ imposing three constraints, and the traceless condition $h^{i}{}_i = 0$ adding one more restriction.

Without loss of generality, we assume GWs to propagate along the $z$-axis. In this case, the perturbation can be written as
\begin{equation}\label{h_decomposition}
    h_{ij}(\eta,\vec{x})=
        \begin{pmatrix}
            h_{+}(\eta,z) & h_{\times}(\eta,z) & 0\\
            h_{\times}(\eta,z) & -h_{+}(\eta,z) & 0\\
            0 & 0 & 0
        \end{pmatrix},
\end{equation}
where $h_+$ and $h_\times$ denote the plus and cross polarization modes, respectively. 

The affine connection is also perturbed as follows 
\begin{equation}
    \Gamma^{\mu}_{\; \alpha  \beta} = \bar{\Gamma}^{\mu}{}_{\alpha\beta}(\eta) +  \delta\Gamma^{\mu}{}_{\alpha\beta}(\eta, \vec{x}),
\end{equation}
where $\bar{\Gamma}^{\mu}{}_{\alpha\beta}$ is the background connection and $\delta\Gamma^{\mu}{}_{\alpha\beta}$ is its linear perturbation. Although the connection itself is not a tensor, its perturbation $\delta\Gamma^{\mu}{}_{\alpha\beta}$ transforms as a tensor. This can be seen directly from how the affine connection transforms under an arbitrary coordinate transformation: it transforms as a tensor plus a non-homogeneous term. Therefore, when taking the difference between two connections (the background $\bar{\Gamma}$ and the total $\Gamma$), the inhomogeneous terms cancel, and the result is a tensor.

The most general perturbation of the affine connection that produces at most second-order differential equations is given by \cite{Tattersall:2017eav,Sulantay:2022sag,Castillo-Felisola:2024atv}
\begin{align}
    \delta\Gamma^{\mu}{}_{\alpha  \beta}&=B_1(\eta) u^{\mu} \gamma_{1 \alpha \beta}+B_2(\eta) \gamma_{2 (\beta}^{\mu}u_{\alpha)}\nonumber\\
    & +B_3(\eta) \partial^{\mu}\gamma_{3 \alpha \beta} +B_4(\eta) \partial_{(\alpha}\gamma^{\mu}_{4 \beta)},\label{Gamma_pert_ansatz}
\end{align}
where the vector $u_{\mu}$ is defined as $u_{\mu}=(a,0,0,0)$, the functions $B_i$ with $i = 1, \ldots, 4$ depend only on the cosmological background, and the tensors $\gamma_{i\alpha\beta}$ with $i = 1, \ldots, 4$ are  transverse-traceless tensor perturbations. These perturbations are defined, in analogy to Eq.~\eqref{h_decomposition}, as follows
\begin{equation}
    \gamma_{i,jk}=
    \begin{pmatrix}
        h_{i+}(\eta,z) & h_{i\times}(\eta,z) & 0\\
        h_{i\times}(\eta,z) & -h_{i+}(\eta,z) & 0\\
        0 & 0 & 0
    \end{pmatrix},
\end{equation}
where $i$ denotes each perturbation matrix $\gamma_{i}$, and the  indices $j,k$ denote the spatial matrix coefficients, while the temporal components of $\gamma_{i\alpha\beta}$ (with either $\alpha=0$ or $\beta=0$) vanish. 

To simplify the subsequent computation, without loss of generality, we set $B_i = 1$ for all $i$. This can always be achieved by a suitable redefinition of the tensor perturbations $\gamma_i$, as shown in  \cite{Tattersall:2017eav,Sulantay:2022sag}. The resulting non-trivial components of the perturbed connection are then explicitly given by:
\begin{widetext}
\begin{equation}
\label{eq:perturbed connection}
  \begin{aligned}
    \delta \Gamma^{0}{}_{11} & =-\frac{h_{1+}}{a} - \frac{1}{a^2}\partial_{\eta}h_{3+}, &
    \delta \Gamma^{0}{}_{12} & =-\frac{h_{1\times}}{a} - \frac{1}{a^2}\partial_{\eta}h_{3\times}, &
    \delta \Gamma^{0}{}_{22} & = \frac{h_{1+}}{a} + \frac{1}{a^2}\partial_{\eta}h_{3+}, \\
    \delta \Gamma^{1}{}_{01} & = \frac{1}{2}\left(\frac{h_{2+}}{a} + \partial_{\eta}\left(\frac{h_{4+}}{a^{2}}\right)\right), &
    \delta \Gamma^{1}{}_{02} & = \frac{1}{2}\left(\frac{h_{2\times}}{a} + \partial_{\eta}\left(\frac{h_{4\times}}{a^{2}}\right)\right), & 
    \delta \Gamma^{1}{}_{13} & = \frac{1}{2}\partial_{z}\left(\frac{h_{4+}}{a^2}\right), &
    \delta \Gamma^{1}{}_{23} & = \frac{1}{2}\partial_{z}\left(\frac{h_{4\times}}{a^2}\right),\\
    \delta \Gamma^{2}{}_{01} & = \frac{1}{2}\left(\frac{h_{2\times}}{a} + \partial_{\eta}\left(\frac{h_{4\times}}{a^{2}}\right)\right), &
    \delta \Gamma^{2}{}_{02} & = -\frac{1}{2}\left(\frac{h_{2+}}{a} + \partial_{\eta}\left(\frac{h_{4+}}{a^{2}}\right)\right), & 
    \delta \Gamma^{2}{}_{13} & = \frac{1}{2}\partial_{z}\left(\frac{h_{4\times}}{a^2}\right), &
    \delta \Gamma^{2}{}_{23} & = \frac{1}{2}\partial_{z}\left(\frac{h_{4+}}{a^2}\right),\\
    \delta \Gamma^{3}{}_{11} & = \partial_{z}\left(\frac{h_{3+}}{a^2}\right), &
    \delta \Gamma^{3}{}_{12} & = \partial_{z}\left(\frac{h_{3\times}}{a^2}\right), &
    \delta \Gamma^{3}{}_{22} & = -\partial_{z}\left(\frac{h_{3+}}{a^2}\right). \\
    \end{aligned}
\end{equation}
\end{widetext}
Note that the connection coefficients $\Gamma^{0}{}_{00}$, $\Gamma^{0}{}_{33}$ and $\Gamma^{3}{}_{30}$ remain purely background functions. In contrast, $\Gamma^{0}{}_{11}$, $\Gamma^{0}{}_{22}$, $\Gamma^{1}{}_{10}$, and $\Gamma^{2}{}_{20}$ consist of a combination of background and perturbed functions. All other components contain purely perturbation terms.

The scalar field $\vartheta$ is also perturbed as follows
\begin{equation}
    \vartheta = \vartheta(\eta) + \delta\vartheta(\eta,\vec{x}),
\end{equation}
where $\vartheta(\eta)$ is the background scalar field and $\delta \vartheta$ is its perturbation. In a cosmological FRW background, scalar perturbations do not couple with tensor perturbations at linear order, which is why $\delta \vartheta$  does not appear in the perturbed tensor equations of motion and can be ignored.

The matter perfect fluid is also perturbed; however, these types of perturbations are only of scalar and vector type, and as a consequence of the SVT decomposition, they do not couple with the tensor perturbations, which are responsible for the propagation of GWs. We thus also ignore matter perturbations.

\subsection{Equations of Motion}

Now that we have the perturbed ansatz for all the fields, we substitute them into the equations of motion from Sec. \ref{sec:general_eqns}. It will be convenient to work in Fourier space, where
\begin{align}
    \gamma_{i}(\eta,z) & = \int \mathrm{d}^{3}k\; \bar{\gamma}_{i}(\eta,\vec{k})e^{i\vec{k}\cdot \vec{x}}, \\
    h(\eta,z) & = \int \mathrm{d}^{3} k\;\bar{h}(\eta,\vec{k})e^{i\vec{k}\cdot \vec{x}}.
\end{align}
This transformation simplifies the analysis by converting the spatial derivatives along the propagation direction into a multiplicative factor of the wavenumber $k$. Furthermore, due to the presence of the CS term, for the perturbations of the metric tensor $h_{ij}$ and the affine connection $\gamma_{i,jk}$, it will be useful to change the polarization basis from the linear modes $+$ and $\times$ to a circular basis Left and Right, defined as follows
\begin{align}
\label{eq: left/right polarization}
    \bar{h}_R(\eta, \vec{k}) & = \frac{\bar{h}_{+} - i\bar{h}_\times }{\sqrt{2}}, &
    \bar{h}_L(\eta, \vec{k}) & = \frac{\bar{h}_{+}+ i\bar{h}_\times }{\sqrt{2}},\\
    \bar{\gamma}_{iR}(\eta, \vec{k}) & = \frac{\bar{\gamma}_{i+} - i\bar{\gamma}_{i\times} }{\sqrt{2}}, &
    \bar{\gamma}_{iL}(\eta, \vec{k}) & = \frac{\bar{\gamma}_{i+} + i\bar{\gamma}_{i\times} }{\sqrt{2}}.
\end{align}
This circular basis is particularly useful since it will decouple the equations of motion for L and R polarizations, which otherwise would mix the linear $+$ and $\times$ polarizations. The perturbed field equations then become\footnote{These equations of motion are written in Fourier space; for the sake of simplicity we drop the notation with bars: $\bar{h} $ and $\bar{\gamma}$.}
\begin{widetext}
\begin{align}
    -2f_R\gamma_{4P}a'^2 + aa'\left(f_R \gamma'_{4P} - f_R' \gamma_{4P}\right) + a^3\left(\gamma_{1P}f_R' + \frac{1}{2}\gamma_{2P} f_R' + 2a'h_{P}f_R' + f_R\gamma'_{1P} + h_{P} f_Ra''\right) \nonumber\\  + \frac{1}{2}a^4h_{P} f''_R + 
    \frac{a^2}{2}\Big(f'_R\left(2\gamma'_{3P} + \gamma'_{4P}\right) + 2f_R\left(k^2\gamma_{3P} + a'\gamma_{1P} + a'\gamma_{2P} + a'^2h_{P} + \gamma''_{3P}\right)\Big)  & = 0, \label{eq: pert metric} \\
    a^4f_R\gamma_{2P}\kappa - a^5f_Rh'_{P}\kappa + a^3f_R\gamma'_{4P}\kappa -2a^2a'f_R\gamma_{4P}\kappa - k \alpha_{P} \vartheta'  \big(2a'\gamma_{4P} - a\gamma'_{4P} - a^2\gamma_{2P}\big) & = 0,
    \label{eq: pert connection 1}\\
    2\kappa a^4f_R\left(2\gamma_{1P} + \gamma_{2P} + 2a'h_{P}\right) + 2\kappa a^5f'_R h_{P} + 2\kappa a^3f_R\big(2\gamma'_{3P} + \gamma'_{4P}\big) - 4 \kappa a^2f_Ra'\gamma_{4P} \nonumber \\
    + k \alpha_{P} \vartheta'\bigg(a'(-4\gamma_{3P} + 2\gamma_{4P}) + \frac{a}{f_R}\left( 4f_R\gamma'_{3P} + f'_R\gamma_{4P} - 2f'_R\gamma_{3P}\right) + 4a^2\gamma_{1P}\bigg) & = 0, \label{eq: pert connection 2} \\
    2k \kappa f_R^2a^4\big(2\gamma_{3P} + \gamma_{4P}\big) + \alpha_{P}\vartheta'\big(4f_Ra'^2\gamma_{4P} - a^3f'_R\gamma_{2P} + 2aa'(f'_R\gamma_{4P} - f_R\gamma'_{4P}) + a^2(4k^2f_R\gamma_{3P} - 2f_Ra'\gamma_{2P}) - f'_R\gamma'_{4P}\big) & = 0, \label{eq: pert connection 3}\\ 
    -2k\kappa a^6f_R^2h_{P} + 2k \kappa a^4f_R^2\gamma_{4P} -\alpha_{P}\vartheta'\big(4a'^2\gamma_{4P} - a^3f'_R\gamma_{2P} - 2aa'(f_R\gamma'_{4P} - f'_R\gamma_{4P}) - a^2(2f_Rk^2\gamma_{4P} + 2f_Ra'\gamma_{2P} + f'_R\gamma'_{4P})\big) & = 0, \label{eq: pert connection 4}
\end{align}
\end{widetext}
where $\alpha_{p} = \alpha \xi_{L,R}$ is an operator whose values are polarization dependent $\xi_L = 1$ and $\xi_R = -1$, and $P=\left(L,R\right)$. The first equation, Eq.~\eqref{eq: pert metric}, corresponds to the perturbed metric, while the remaining four equations are associated with the perturbed affine connection. 

It is worth noting that the perturbed metric equation preserves parity symmetry (i.e.\ it does not depend on $\alpha$), whereas the perturbed affine equations break it due to the contribution from the Chern-Simons term. This asymmetry can be identified through the coefficients $\alpha_{L,R}$, which distinguish the Left and Right-handed polarization modes by an opposite sign.

While we initially have five tensor perturbations, the system can be reduced to a differential equation for a single physical tensor degree of freedom. This is obtained by first solving the perturbed affine equations: one begins by determining $\gamma_{2P}$ from Eq.~\eqref{eq: pert connection 1} and subsequently uses this result to obtain $\gamma_{1P}$ from Eq.~\eqref{eq: pert connection 2}. With the connection perturbations $\gamma_{1P}$ and $\gamma_{2P}$ known, one can then obtain expressions for $\gamma_{4P}$ and $\gamma_{3P}$ from Eqs.~\eqref{eq: pert connection 3} and \eqref{eq: pert connection 4}, respectively. The complete expressions for the perturbed affine solutions are provided in Appendix~\ref{sec:appendix A, solution}. Substituting these results into the perturbed metric equation yields the following propagation equation for GWs 
\begin{align}
\label{eq: gw vp}
    &h''_{P} + \left(\frac{B_{mn}\alpha_{P}^{m}k^{n}}{A_{mn}\alpha_{P}^{m}k^{n}}\right)h'_{P} 
    +  \left(\frac{C_{mn}\alpha_{P}^{m}k^{n}}{D_{mn}\alpha_{P}^{m}k^{n}}\right)h_{P} = 0,
\end{align}
where the coefficients $A_{mn}$, $B_{mn}$, $C_{mn}$, and $D_{mn}$ are defined completely in terms of background quantities, and a summation over the indices $m$ and $n$ is implicit. Explicit expressions are given in Appendix \ref{sec:appendix B, series expansion}. It is worth noting that we find some non-vanishing coefficients with $m = 0$, which means that even in the limit when $\alpha=0$, the coefficients $A_{0n}\not=0$, $B_{0n}\not=0$, and $C_{0n}\not = 0$  receive non-trivial contributions from the $f(R)$ interaction. In this regime, the theory reduces to Palatini $f(R)$ gravity, where parity symmetry is restored, yet modifications in the GW propagation persist.

Eq.~\eqref{eq: gw vp} represents the general propagation equation for GWs in an arbitrary $f(R)$ theory of gravity coupled to a dCS term and a scalar field. For odd values of $m$, the terms $\alpha_P^m$ take different values for the L and R polarization, leading to birefringence. The modifications in the friction term $h'_P$ lead to amplitude birefringence (that is, relative changes in the amplitude of L and R polarizations) and those in the $k^2 h_P$ term lead to velocity birefringence (relative changes in the phase and velocity of L and R polarizations).

It is instructive to verify the consistency of this result in specific limits. By setting $f(R) = R$ and switching off the dCS contribution, Eq.~\eqref{eq: gw vp} reduces to the standard GW propagation equation of GR. Conversely, keeping $f(R) = R$ while retaining the dCS contribution reproduces the well-known GW equation in dCS gravity \cite{Peng:2022ttg,Sulantay:2022sag}. Our result therefore consistently recovers the expected GR and GR + dCS limits, while at the same time generalizing them to an arbitrary function $f(R)$. 

Finally, it is instructive to compare these results with those of the metric formalism. In that case, there is no independent connection perturbation, and the system reduces to a single perturbed metric equation from the outset. As a consequence, the metric formalism yields a GW propagation equation that contains only amplitude birefringence, while velocity birefringence is absent. The resulting propagation equation takes the simpler form
\begin{equation}
    h''_{P} + \left(\frac{2\mathcal{H} + \frac{f'_R}{f_R} + \frac{2\alpha_{L,R} k\vartheta''}{a^2 \kappa f_R}}{1 + \frac{2\alpha_{L,R} k\vartheta'}{a^2\kappa f_R}}\right)h'_{P} + k^2h_{P} = 0,
\end{equation}
which is in agreement with previous results \cite{Alexander:2007kv, Yunes:2010yf, Yagi:2017zhb, Sulantay:2022sag} when choosing $f(R) = R$.
By contrast, the Palatini formalism introduces additional complexity through the independent connection perturbations, which are responsible for the velocity birefringence in Eq.~\eqref{eq: gw vp}.

The counting of the total degrees of freedom of this model requires an extensive Hamiltonian analysis \cite{Banados:1996yj,Banados:2025sww}. Nonetheless, in the current framework (cosmological scenario), the model contains the usual spin-2 tensor modes, with their respective speeds shown in the next Section. Establishing the remaining degrees of freedom associated with spin-0 (scalar) and spin-1 (vector) is a task beyond the scope of our analysis.

\section{Phenomenology}
\label{sec:phenomenology}

The GW propagation equation \eqref{eq: gw vp}  can be written in a more compact manner as follows, making the polarization explicit,
\begin{equation}
\label{eq: gws dcs}
    h''_{L,R} + \Xi_{L,R}h'_{L,R} + \omega^2_{L,R}  h_{L,R} = 0,
\end{equation}
where $\Xi_{L,R}$ is a friction/dissipation term and $\omega^2_{L,R}$ is the angular velocity of the GW, defined as
\begin{align}
    \Xi_{L,R} & = \frac{ B_{mn}\alpha_{L,R}^{m}k^{n}}{ A_{mn}\alpha_{L,R}^{m}k^{n}}, & 
    \omega^2_{L,R} & = \frac{C_{mn}\alpha_{L,R}^{m}k^{n}}{D_{mn}\alpha_{L,R}^{m}k^{n}}.
\end{align}
Both quantities are polarization-dependent: for the L and R polarizations, the dissipation and frequency terms differ, which gives rise to amplitude and velocity birefringence, respectively. For comparison, in the metric formalism of dCS, only amplitude birefringence appears; therefore, velocity birefringence is a direct consequence of the Palatini formalism. 

The general solution to Eq.~\eqref{eq: gws dcs} is found by implementing a WKB approximation, where we assume the period of the GW to be much shorter than the cosmological timescale (the standard adiabatic condition $k \gg \mathcal{H}$). The solution then takes the form \cite{Sulantay:2022sag}
\begin{equation}
\label{solution gw modified equation}
    h_{L,R}(\eta,k) \approx A_{L,R}e^{-\int \Xi_{L,R}(\tilde{\eta},k)\mathrm{d}\tilde{\eta}}e^{\pm i\int \Omega_{L,R}(\tilde{\eta},k)\mathrm{d}\tilde{\eta}},
\end{equation}
where the integration limits are from time of emission to time of detection, implying that the observed signal encodes the accumulated effect of birefringence over its propagation history. The function $\Omega_{L,R}$ is defined as
\begin{equation}
\label{Omega LR}
    \Omega^2_{L,R} = \omega^2_{L,R} - \Xi^2_{L,R}.
\end{equation}
Note that in order to have an oscillatory solution, we require $\Omega^2_{L,R} > 0$, which imposes the condition $\omega_{L,R} \gg \Xi_{L,R}$. 

Eq.~\eqref{solution gw modified equation} can alternatively be written in a more illustrative form as
\begin{equation}\label{eq:mus}
    h_{L,R}(\eta,k) \approx h^{GR}_{L,R}(\eta,k)e^{-\mu_A(\eta,k)}e^{\pm \mu_V(\eta,k)},
\end{equation}
where $h^{GR}_{L,R}$ is the standard GR solution, and $\mu_A$ (dimensionless) and $\mu_V$ (units of radians) characterize the strength of amplitude and velocity birefringence, respectively. Both quantities will be discussed later in this section.

In order to understand the behavior of GWs, it is useful to Taylor-expand $\Xi_{L,R}$ and $\omega_{L,R}$ for a small CS coupling $\alpha$. For the angular velocity we obtain
\begin{widetext}
\begin{align}
\label{gw: frequency}
    \omega^2_{L,R} & \approx k^2 + \frac{3k^2\alpha^2_{L,R}\vartheta'}{4\kappa^2a^6f_R^4}\left(5a^2\vartheta'(\partial_{\eta}f_R)^2 + 
    4f_R^2\left(4a'^2\vartheta'-a\vartheta'a''-2aa'\vartheta''\right) -
    2af_R\left(-6a'\vartheta'\partial_{\eta}f_R + a\left(\vartheta\partial_{\eta\eta}^2f_R + 2\vartheta''\partial_{\eta}f_R\right)\right)\right) \nonumber\\
    & +  \frac{k^3\alpha_{L,R}^3\vartheta'^2}{2\kappa^3 a^8 f_R^5}\left(-11a^2\vartheta'\partial_\eta f_R^2 + 2f_R^2(-21a'^2\vartheta + a\vartheta' a'' + 9 aa'\vartheta'') + af_R(-40a'\vartheta'\partial_\eta f_R + a(\vartheta\partial_{\eta\eta}f_R + 9\vartheta''\partial_\eta f_R))\right),
\end{align}
\end{widetext}
where for clarity we have made explicit the conformal-time derivatives. We see that parity-preserving and parity-violating deviations from GR appear simultaneously, with parity-breaking terms in the velocity becoming relevant at third order in $k\alpha$. Due to the $k$-dependence of this term, the dispersion relation of GWs is modified, which can significantly distort the phase evolution of a GW signal.

Similarly, the friction term can be expanded as
\begin{align}
\label{gw: dissipation}
    \Xi_{L,R} \approx \left(2\mathcal{H} + \frac{f'_R}{f_R} \right) 
    + \frac{2k\alpha_{L,R}}{\kappa a^3f_R^2}\left(a\vartheta'f'_R + f_R\left(2a'\vartheta' - a\vartheta''\right)\right),
\end{align}
where the first term contains two contributions: the standard GR contribution $2\mathcal{H} = \Xi_{GR}$ and a correction coming from the derivatives of $f(R)$. The second term depends on background quantities, such as the scale factor $a(\eta)$, the scalar field $\vartheta(\eta)$, and $f(R)$ and contains a linear contribution in the CS coupling constant $\alpha_{L,R}$, which is responsible for breaking parity symmetry. Eqs.\ (\ref{gw: frequency})-(\ref{gw: dissipation}) generalize the expressions already found in \cite{Sulantay:2022sag}.

Note that by fixing $f(R) = R$ and switching off the dCS contribution, $\alpha_{L,R} \to 0$ we recover the GR expressions for the frequency and friction terms. Similarly, by fixing $f(R) = R$ and keeping the dCS contribution, we recover the results from GR coupled to dCS found in \cite{Sulantay:2022sag}.

\begin{figure*}[t]
    \centering
    \includegraphics[width=\textwidth]{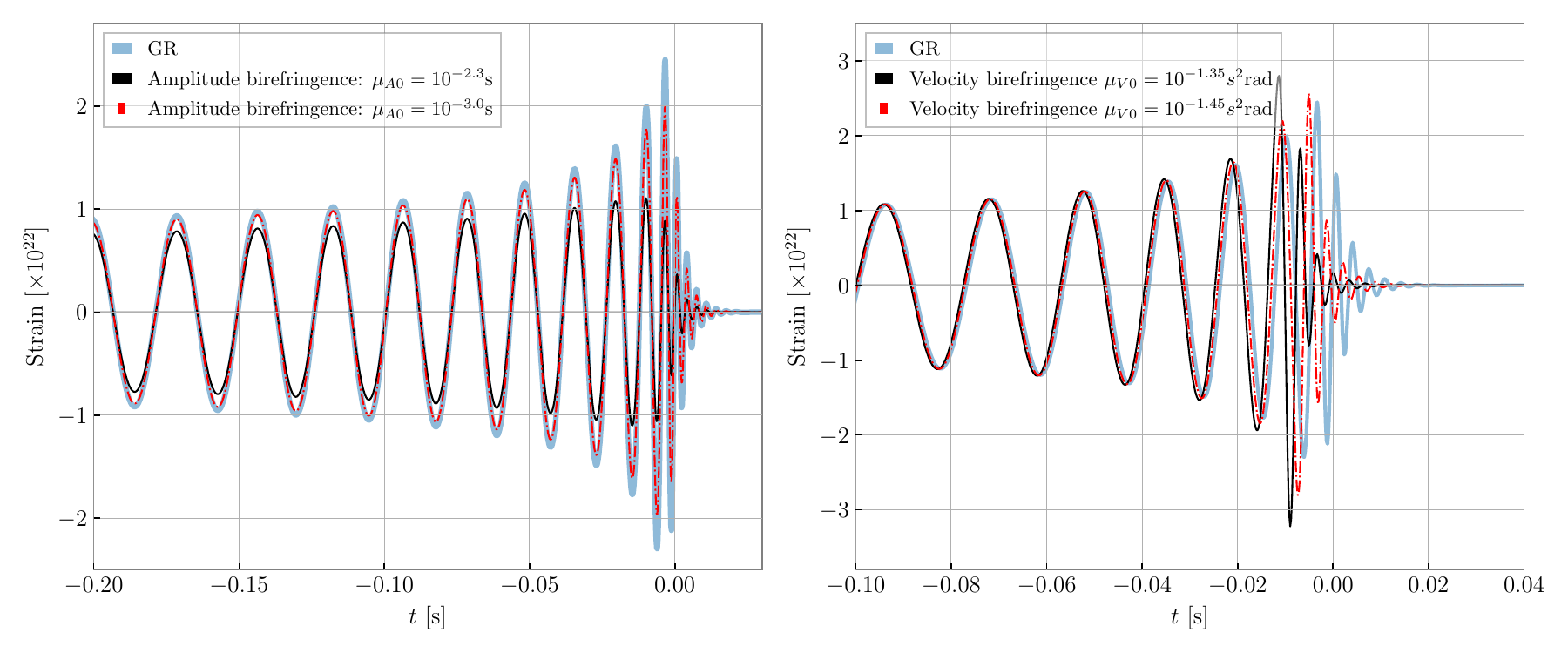}
    \caption{Illustration of amplitude (left) and velocity (right) birefringence for a GW signal with a purely left-handed polarization. In blue, it is shown the standard prediction of the strain of the GW in GR, while black and red show two examples of birefringence. Note that, the velocity birefringence effect shown here is predicted by Chern-Simons in the Palatini formalism but not in the metric formalism.}
    \label{fig:avb}
\end{figure*}

The friction $\Xi_{L,R}$ and $\omega^2_{L,R}$ depend on power-law contributions in the wavenumber $k$, which arise from Fourier transforming spatial-derivative terms; the model thus exhibits only higher-order spatial derivatives, while higher-order temporal derivatives are absent. Therefore, Ostrogradsky ghosts, which are associated with higher-order time derivatives and generically lead to a Hamiltonian that is unbounded below, are not present in this model.

Regarding Laplacian instabilities, the highest power of $k$ appearing in the numerator of $\omega^2_{L,R}$ comes from the coefficients $C_{24}$ and $C_{44}$, both of order $k^4$, while the highest power of $k$ in the denominator is $D_{22}$, of order $k^2$ (see Appendix B for their explicit expressions). The coefficient $C_{24}$ is manifestly positive definite, and, since $\alpha_{L,R}$ is a small parameter, the term $C_{24}\alpha^2_{L,R} \gg C_{44}\alpha^4_{L,R}$ within the regime of validity of the small-coupling expansion, so the leading, dominant contribution to the numerator remains positive. Similarly, $D_{22}$ is positive definite and constitutes the dominant term in the denominator. As both leading terms are parity-invariant and positive, $\omega^2_{L,R}$ remains positive at large $k$, and no Laplacian instability arises within the regime of validity of the small-coupling expansion adopted throughout this work.

The parameters characterizing the strength of the modifications beyond GR are $\mu_{A}$ (amplitude birefringence) and $\mu_V$ (velocity birefringence); the former is defined as
\begin{align}
    \mu_A  = \int \Delta\Xi_{L,R}(\tilde{\eta},k)\mathrm{d}\eta 
\end{align}
where $\Delta \Xi(\tilde{\eta},k) =  \Xi_{L,R} - \Xi_{GR}$ quantifies the deviation from GR. According to Eq.\ (\ref{gw: dissipation}), the leading-order correction is linear in the CS coupling constant $\alpha_{L,R}$ and linear in $k$, which translates into a linear dependence on the frequency $f$ of the GW. 

For a quasi-circular, nearly equal-mass, non-precessing binary, the inclination angle $\iota$ between the line of sight and the binary angular momentum vector determines the emitted amplitude of each circular polarization mode. In particular, for $\iota = 0$, the Right polarization mode vanishes, leaving only the Left mode; this results in a decrease in the amplitude of the GW if $\mu_A > 0$, and an increase otherwise. Conversely, if $\iota = \pi$, the Left mode vanishes, leaving only the Right mode, and the GW amplitude increases if $\mu_A < 0$; otherwise, it decreases.

The velocity birefringence parameter $\mu_{V}$ is defined as
\begin{align}
        \mu_V  = \int \Delta\Omega_{L,R}(\tilde{\eta})\mathrm{d}\tilde{\eta},
\end{align}
where $\Delta \Omega_{L,R}(\tilde{\eta}) = \Omega_{L,R} - \Omega_{GR}$, which can be written as 
\begin{equation}
    \mu_V  = \int \left(\sqrt{\omega^2_{L,R} - \Xi^2_{L,R}} - \sqrt{\omega^2_{GR} - 4\mathcal{H}^2} \right)\mathrm{d}\tilde{\eta},
\end{equation}
where $\omega^2_{GR} = k^2$. Note that the first contribution  in $\omega^2_{L,R}$ coming from dCS is $\left(k\alpha_{L,R}\right)^2$, which is parity invariant. Therefore, one needs to go to higher order to obtain the leading parity-violating term, of order $\left(k\alpha_{L,R}\right)^3$. Although this contribution appears as $k^3$ in $\omega^2$, it will appear as $\mu_V \sim k^2$ in a small $\alpha$ expansion due to the square root. This will thus lead to frequency-dependent changes in the phase of a GW signal.

We can write a modified dispersion relation given by:
\begin{equation}
    \omega^2_{L,R} \approx k^2\left(1 + \alpha^2_{L,R}\delta_1 + \alpha^3_{L,R} k \delta_2\right),
\end{equation}
where $\delta_1$ and $\delta_2$ are functions of background quantities only, namely the scale factor $a(\eta)$, the scalar field $\vartheta(\eta)$, and the function $f(R)$. A Taylor expansion in $\epsilon = \alpha^2_{L,R}\delta_1 + \alpha^3_{L,R} k \delta_2 \ll 1$ yields the following group velocity
\begin{equation}\label{eq:group_velocity}
    v_{g L,R} = \frac{d \omega_{L,R}}{d k} \approx 1 + \alpha_{L,R}^2\frac{\delta_1}{2} + k \alpha_{L,R}^3\delta_2.
\end{equation}
The first-order correction $\delta_1$ introduces a shift in the global propagation speed of GWs, whereas the higher-order contribution, which breaks parity symmetry, introduces a frequency dependence in the group velocity of the GW, and therefore can compress or stretch the waveform. 

In order to illustrate the effect of amplitude and velocity birefringence on a GW signal, let us consider the strain response for a LIGO-type detector, $h_{strain} = F_p h_+ + F_\times h_\times$, with $F_p$ and $F_\times$ being the antenna patterns for the plus and cross polarizations, defined as
\begin{align}
    F_+ & = \frac{1}{2}\left(1 + \cos^2\theta\right)\cos 2\phi \cos2\psi - \cos\theta\sin2\phi\sin 2\psi, \\
    F_\times & = \frac{1}{2}\left(1 + \cos^2\theta\right)\cos 2\phi \sin2\psi - \cos\theta\sin2\phi\cos 2\psi,
\end{align}
where we followed the notation of \cite{Lagos:2024boe}. The angles $(\theta,\phi)$ are polar and azimuthal angles that determine the sky location of the source relative to the detector, and $\psi$ is the orientation angle that defines the orientation of the principal axes of the linear polarization frame in the plane of the sky.

Figure \ref{fig:avb} shows the effect of amplitude (left panel) and velocity (right panel) birefringence, for two different examples and a comparison to GR. We show the last oscillations before the merger of a stellar-mass binary black hole, with equal masses $m_1 = m_2 = 30 M_\odot$, redshift $z=0.5$, and inclination angle $\iota = 0$. For illustration purposes, we define $\mu_A = \mu_{A0} k$, and $\mu_V = \mu_{V0}k^2$, where $k$ is the wavenumber of the GW \footnote{Here we are considering the leading $k$-dependent term, but in practice $\mu_{V}$ also contains $k^{0}$ terms, which have been studied in \cite{Lagos:2024boe}.}.

In the left panel of Fig.\ \ref{fig:avb} we show the prediction of GR in blue, and the predictions including amplitude birefringence introduced by the dCS term in red and black. For $\iota = 0$, the Right polarization mode vanishes completely, leaving only the Left mode. For $\mu_{A0} = 10^{-3.0} \si{s} > 0$, the amplitude is decreased, and for the value $\mu_{A0} = 10^{-2.3} \si{s}$, amplitude birefringence is stronger, yielding a considerable suppression of the strain of the GW. Since amplitude birefringence is proportional to the frequency, the amplitude suppression is stronger near the merger than during the inspiral.

In the right panel of Fig.\ \ref{fig:avb} we show an analogous plot for velocity birefringence. In this case, there is a change in the phase of the signal, which compresses the GW signal duration. For the case $\iota = 0$ shown in Fig.\ \ref{fig:avb}, the Right polarization vanishes, leaving only the Left mode. In the case of $\mu_V>0$ (as in the plot), higher-frequency modes associated with the merger propagate faster than lower-frequency modes from the inspiral (see Eq.\ (\ref{eq:group_velocity})); thus the waveform is compressed. Conversely, if $\iota = \pi$, only the Right mode survives and the signal is stretched instead.  

Finally, we compare these results with the metric formalism. In this case, the GW equation reads:
\begin{equation}
\label{gw equation in metric formalism}
    h''_{L,R} + \Xi_{L,R}h'_{L,R} + \omega^2h_{L,R} = 0,
\end{equation}
where $\Xi_{L,R}$ is the friction term of the GW, defined in Appendix~\ref{sec:appendix C, metric formalism} and $\omega^2 = k^2$. Comparing  the GW equation in the metric formalism, Eq.\ \eqref{gw equation in metric formalism}, with that in the Palatini formalism, Eq.~\eqref{eq: gws dcs}, we see the absence of velocity birefringence; therefore, no stretching or squeezing of the waveform is expected, nor are there modifications to the GW propagation speed. In the metric formalism, only amplitude birefringence is present.

Additionally, the solutions in the Palatini formalism, presented in Eqs.~\eqref{solution gw modified equation} and \eqref{eq:mus}, are also valid in the metric formalism. As in the Palatini formalism, if we consider  $\alpha_{L,R}$ to be a small parameter, the Taylor expanded expression for the friction term agrees with that of the Palatini formalism, Eq.~\eqref{gw: dissipation}, to leading order in $\alpha_{L,R}$.

\subsection{Late-time background universe: de-Sitter}

We are interested in exploring the phenomenology of this model in the late-time universe, where the accelerated expansion arises from the $f(R)$ term rather than from a cosmological constant or a minimally coupled scalar field. 

As an approximation, we analyze a de-Sitter universe, characterized by positive, constant scalar curvature $R$. Under this assumption, the angular velocity $\omega^2_{L,R}$ simplifies considerably to
\begin{align}
\label{eq: de-sitter omega}
    \omega^2_{L,R} & \approx k^2 + \frac{3k^2\alpha^2_{L,R}\vartheta'}{\kappa^2a^6f_R^2}
    \left(4a'^2\vartheta'-a\vartheta'a''-2aa'\vartheta''\right) \nonumber\\
    & +  \frac{k^3\alpha_{L,R}^3\vartheta'^2}{\kappa^3 a^8 f_R^3}\left( -21a'^2\vartheta + a\vartheta' a'' + 9 aa'\vartheta'' \right),
\end{align}
where all the time (conformal) derivatives of $f(R)$ vanish. Note that the above expression is similar to the one already found in the literature \cite{Sulantay:2022sag}. Nonetheless, the introduction of $f(R)$ couples to the dCS contribution and can thus suppress (if $f_R>1$) or enhance velocity birefringence (if $0<f_R<1$). 

Moreover, the friction term $\Xi_{L,R}$ in a de-Sitter type of universe simplifies to
\begin{equation}
\label{eq: de-sitter friction}
    \Xi_{L,R} \approx 2\mathcal{H}  + \frac{2k\alpha_{L,R}}{\kappa a^3f_R}\left(2a'\vartheta' - a\vartheta''\right).
\end{equation}
This expression again matches the results of previous works \cite{Sulantay:2022sag}, up to an enhancement or suppression of the birefringence effect due to the presence of the function $f(R)$.

Therefore, at late times, the effect of an $f(R)$  theory of gravity is to increase or reduce the amplitude and velocity birefringence effects (in both the metric and Palatini formalisms), as can be seen directly from Eqs.~\eqref{eq: de-sitter friction} and \eqref{eq: de-sitter omega}. Whether an enhancement or suppression of birefringence occurs will depend on the value of $f_R$, which in turn depends on the specific functional form of the $f(R)$ function assumed. 

In order to provide an accurate description of the phenomenological predictions of $f(R)$ gravity, it is necessary to carefully choose the functional form of $f(R)$. The chosen $f(R)$ must satisfy several conditions to ensure that the model provides a realistic description of dark energy, in accordance with cosmological survey data. The main conditions are: (i) positivity of the effective gravitational constant, which requires that the derivative $f_R = \partial_R f(R)$ be positive \cite{Santos:2012vs,Campista:2010jb}; (ii) the model should asymptotically tend to $\Lambda$CDM, while at the same time the modifications coming from $f(R)$ should vanish in the early universe, since dark energy is negligible at high redshifts. These two conditions must be satisfied in both the Palatini and metric formalisms.

Next, we present some examples of $f(R)$ functions considered in the literature and examine whether they predict enhancement ($0 < f_R < 1$) or suppression ($f_R > 1$) of the birefringence effects. For each example, we calculate the scalar curvature today, $R_0=R(z=0)$, and the value of $f_{R_0}=f_R(R_0)$ today.

\subsubsection{Palatini formalism}

In order to compute the scalar curvature today, $z = 0$, we start from the trace equation, which in this formalism is an algebraic equation for $R$. The trace equation takes the form of Eq.~\eqref{eq: trace palatini}, namely
\begin{equation}
\label{eq: trace palatini matter}
    Rf_R(R) - 2f(R) = -3H_0^2\Omega_{m0}(1 + z)^3,
\end{equation}
where the matter contribution to the trace of the energy-momentum tensor includes only the non-relativistic matter component, whereas the scalar field contribution is neglected, assuming $\Omega_{\vartheta 0} (z = 0) \ll 1$, since the accelerated expansion of the universe is driven by $f(R)$. Note that this assumption implies that the dCS scalar field plays a negligible role in the background dynamics, however, this does not imply that its contribution to the birefringence can be neglected, because the strength of the dCS and the scalar field are characterized by the independent coupling constants $\alpha$ and $\beta$, respectively. When a given $f(R)$ function is specified, evaluating the above expression at $z = 0$ yields the value of $R_0$.

\paragraph{Hu-Sawicki model.}
The Hu-Sawicki model \cite{Hu:2007nk} describes the accelerated expansion of the universe, where the \textit{cosmological constant} is an effective, emergent quantity arising from the $f(R)$ function. It is defined as follows 
\begin{equation}
    f_{HS}(R) =  R - m^2\frac{c_1\left(\frac{R}{m^2}\right)^{n}}{c_2\left(\frac{R}{m^2}\right)^{n} + 1},
\end{equation}
where $c_1$ and $c_2$ are dimensionless parameters, $n$ is a positive integer and $m^2 = H_0^2 \Omega_{m0}$, with $H_0$ being the current value of the Hubble constant and $\Omega_{m0}$ the present matter density parameter. For $n = 1$, the case considered in this work, the $f_{HS}(R)$ reduces to
\begin{equation}
    f_{HS}(R) = R - 2\frac{\Lambda}{1 + \frac{\mu^2}{R}},
\end{equation}
where $\Lambda = m^2c_1/c_2$ and $\mu^2 = m^2/c_2$. In the high-curvature regime, $R \gg \mu^2$, the model resembles the $\Lambda$CDM model. The first derivative of $f(R)$ gives
\begin{equation}
    f_R(R_0) =\frac{3 + 2 \left(\frac{R_0}{m^2}\right)^2c_2}{\frac{R_0}{m^2}\left(1 + 2\frac{R_0c_2}{m^2}\right)} > 0 ,
\end{equation}
from which one finds the bounds for the parameter $c_2$. Details can be found in Ref. \cite{Santos:2012vs,Ravi:2024jwl}.

Using observational constraints coming from different combinations of data, including the Supernovae Ia (SNIa), cosmic chronometers (CC), starburst galaxies (H$_{II}$G), and cosmic microwave background (CMB), it is possible to find the best-fit values for the parameters of the Hu-Sawicki model \cite{Ravi:2024jwl}, and hence the value of $f_{R_0}$, as shown in the upper rows of Table \ref{table: HS + E}.
\begin{table}[h]
    \centering
    \begin{tabular}{|c|c|c|} \hline
         Model & Data &  $f_{R_0}$ \\ \hline
         Hu-Sawicki & SNIa + CC & $0.91$\\
         & SNIa + CC + H$_{II}$G & $0.90$ \\
         & SNIa + BAO + CC + H$_{II}$G + CMB & $0.96$ \\ \hline
         Exponential & SNIa + CC & $0.99$\\
         & SNIa + CC + H$_{II}$G & $0.99$ \\
         & SNIa + BAO + CC + H$_{II}$G + CMB & $0.99$ \\ \hline
         Hyperbolic & SNIa + CC & $0.95$\\
         & SNIa + CC + H$_{II}$G & $0.94$ \\
         & SNIa + BAO + CC + H$_{II}$G + CMB & $0.99$ \\ \hline
    \end{tabular}
    \caption{Hu-Sawicki, Exponential, and Hyperbolic gravity best-fit values for $f_{R0}$ in the Palatini formalism \cite{Ravi:2024jwl}.}
    \label{table: HS + E}
\end{table}

From Table \ref{table: HS + E}, all data predict $f_{R_0} <1$, and therefore an enhancement of both amplitude and velocity birefringence effects. The strongest amplification is found using SNIa + CC + H$_{II}G$ data, giving $f_{R_0} \approx 0.90 $, whereas the smallest deviation from $\Lambda$CDM is obtained with SNIa + BAO + CC + H$_{II}$G + CMB, giving $f_{R_0} \approx 0.96$.

\paragraph{Exponential gravity model.}
The exponential gravity model \cite{Cognola:2007zu,Linder:2009jz,Bamba:2010ws,Yang:2010xq} is defined as 
\begin{equation}
    f_E(R) = R + \xi\left(e^{-\sigma R} - 1\right),
\end{equation}
where $\xi > 0$ and $\sigma> 0$ are the free parameters of the model. In the limit $\sigma \to \infty$ with $\xi = 2\Lambda$ , the model reduces to $\Lambda$CDM. Introducing the parametrization
\begin{equation}
    f_E(R) = R - 2\Lambda\left(1 -  e^{-\frac{R}{b\Lambda}}\right),
\end{equation}
where $\Lambda = \xi/2$ and $b = 2/(\xi\sigma)$, the model asymptotically tends to $\Lambda$CDM for finite values of its parameters. In particular, when $\sigma \to 0$, so that $R \gg \sigma\Lambda$,  one recovers $f_E(R) \to R - 2\Lambda$. The condition $f_{R} > 0$ is satisfied as shown in \cite{Bamba:2010ws}.

Using the same observational data, we obtain numerical values for the free parameters of the model and, therefore, compute $f_{R_0}$ \cite{Ravi:2024jwl}. For all data combinations, the Exponential model predicts $f_{R_0} \approx 0.99$, implying only a marginal amplification of both amplitude and velocity birefringence. This represents the smallest deviation from $\Lambda$CDM.

\paragraph{Hyperbolic gravity model.}
Having introduced the Hu-Sawicki and the Exponential gravity models above, we now briefly introduce the Hyperbolic gravity model \cite{Cognola:2007zu}, for which $f(R)$ is defined as
\begin{equation}
    f_T(R) = R - \xi R_T \tanh\left(\frac{R}{R_T}\right),
\end{equation}
where $\xi > 0$ and $R_T > 0$ are the free parameters of the model. An alternative parametrization gives
\begin{equation}
    f_T(R) = R - 2\Lambda \tanh\left(\frac{R}{b\Lambda}\right),
\end{equation}
where $\Lambda = \xi R_T/2$ and $b = 2/\xi$. It is straightforward to verify that as $b \to 0$, i.e., $\xi \to \infty$ and $R_T \to 0$, while keeping $\xi R_T$ finite, the model asymptotically approaches to $\Lambda$CDM. The viability condition $f_R > 0$ is well satisfied; see \cite{Cognola:2007zu,Ravi:2024jwl}.

Like the other two models, the Hyperbolic gravity model predicts only an enhancement of both amplitude and velocity birefringence. Its deviation from $\Lambda$CDM is smaller than that of the Hu-Sawicki model but larger than that of the Exponential gravity model.

It is worth emphasizing that the above results were found using the Palatini formalism, where the metric tensor and the affine connection are independent fields, as considered in this paper. Furthermore, the minimally coupled scalar field  has been neglected, consistently with the assumption $\Omega_{\vartheta0} (z=0) \ll 1$ introduced above. In order to incorporate more recent data, such as DESI \cite{Plaza:2025gcv}, we also present results within the metric formalism.

\subsubsection{Metric formalism} 

When working in the metric formalism, it is necessary to impose stability conditions under perturbations. In order to avoid Dolgov-Kawasaki instabilities,  one requires $f_{RR} = \partial_{RR}f(R) > 0$, which is a condition absent in the Palatini formalism, since there is no scalaron field\footnote{A detailed discussion of whether the condition $f_{RR} = \partial_{RR}f(R) > 0$ is required in the Palatini formalism is found in \cite{DeFelice:2010aj}.}, and $f(R)$ does not represent an extra degree of freedom. A detailed discussion of the viability requirements in both formalisms is found in Refs. \cite{Tsujikawa:2007xu,DeFelice:2010aj,Amendola:2015ksp,Plaza:2025gcv}.

Unlike in the Palatini formalism, the trace equation in the metric formalism provides a dynamical evolution for the scalar curvature $R$ (see Appendix~\ref{sec:appendix C, metric formalism}), which reads
\begin{equation}
    Rf_R(R) -2f(R) + 3\square f_R(R) = \kappa T.
\end{equation}
Working in an FRW background, and using $f'_R = f_{RR}R'$, and $f''_{R} = f_{RRR}R'^2 + f_{RR}R''$, the above expression yields
\begin{equation}
    f_RR = -3H_0^2\Omega_{m0}(1 + z)^3 - \frac{3}{a^2}\left(f_{RRR}R'^2 + f_{RR}R'' + 3Hf_{RR}R'\right) + 2f,
\end{equation}
where we have used $T = - \rho_{m}$, neglected the scalar field contribution, $\Omega_{\vartheta 0}(z=0) \ll 1$ consistently with the assumption adopted in the Palatini case, since the accelerated expansion is driven by $f(R)$, and written the non-relativistic matter density as a function of the redshift via the matter continuity equation. Evaluating this expression at $z=0$ and requiring a de-Sitter solution, $R \approx R_{0}$, one obtains
\begin{equation}
    R_0 f_{R_0} \approx 2f(R_0) - 3H_0^2\Omega_{m0},
\end{equation}
which is an algebraic equation for $R_0$, in the de-Sitter limit. Notably, this expression coincides with that obtained in the Palatini formalism, see Eq.~\eqref{eq: trace palatini matter}.

Using the recent observational results from DESI, including cosmic chronometers (CC), SNIa, and BAO, it is possible to find numerical values for the $f(R)$ models of Hu-Sawicki, Exponential, and Hyperbolic gravity \cite{Cognola:2007zu}; details on how to obtain the best-fit values are found in \cite{Plaza:2025gcv}. 

Table \ref{tab:HS/Ex metric formalism} shows the values of $f_{R_0}$ for the different data combinations analyzed in \cite{Plaza:2025gcv}.
\begin{table}[h!]
    \centering
    \begin{tabular}{|c|c|c|}\hline
        Model & Data & $f_{R_0}$ \\ \hline
        Hu-Sawicki & CC + PPS & $0.92$\\ 
        & CC + PPS + BAO$_{1}$ & $0.99$\\ 
        & CC + PPS + BAO$_{2}$ & $0.99$\\ \hline
        Exponential & CC + PPS & $0.69$\\
        & CC + PPS + BAO$_{1}$ & $0.92$\\ 
        & CC + PPS + BAO$_{2}$ & $0.96$\\ \hline
        Hyperbolic & CC + PPS & $0.51$\\ 
        & CC + PPS + BAO$_{1}$ & $0.51$\\ 
        & CC + PPS + BAO$_{2}$ & $0.96$\\ \hline
    \end{tabular}
    \caption{Hu-Sawicki, Exponential, and Hyperbolic gravity best-fit values for $f_{R_0}$ in the metric formalism \cite{Plaza:2025gcv}.}
    \label{tab:HS/Ex metric formalism}
\end{table}

For the Hu-Sawicki model, all data combinations predict $f_{R_0} < 1$, so only an enhancement of the amplitude birefringence is possible. Moreover, as the data set is extended from CC to CC + PPS and then to CC + PPS + BAO$_{1,2}$, $f_{R_0} \sim 0.99$, indicating that the model produces only a small deviation from $\Lambda$CDM. For the Exponential gravity model, the CC + Pantheon$^+$ + SH$0$ES (PPS) data predict a stronger amplification of the birefringence effect; however, complementing these data with BAO yields a weaker amplification, leading to a result similar to that of the Hu-Sawicki model. Finally, for the Hyperbolic model, the CC + PPS data produce the largest amplification of the birefringence effect, with $f_{R_0} \approx 0.51$, representing a considerable deviation from the standard cosmological model. This remains the case when including CC + PPS + BAO$_{1}$; however, with CC + PPS + BAO$_{2}$, $f_{R_0} \approx 0.96$, which is very close to the $\Lambda$CDM. 

It is worth noting that computing $R_0$ requires several assumptions; most importantly, the derivatives of $f(R)$ are neglected in the de-Sitter limit. Therefore, the present analysis is simplified, and a more complete description should account for contributions from $f_R$, $f_{RR}$, and possible contributions from the minimally coupled scalar field $\vartheta$.

\subsection{Gravitational Wave Constraints}
In this section, we discuss observational constraints on gravitational waves and how they apply to the model analyzed in this paper.  

Current observations from the binary neutron star merger GW170817 \cite{LIGOScientific:2017zic} and its EM counterparts impose tight constraints on the propagation speed of GWs, $| c_T - 1| < 10^{-15}$ \cite{LIGOScientific:2017zic}. Considering a dark-energy-dominated universe, with $R \approx  R_0$ approximated as a constant, the angular velocity $\omega^2_{L,R}$ takes the form of Eq.~\eqref{eq: de-sitter omega}, from which one obtains the following constraint on the global propagation speed of GWs:
\begin{equation}
    |c_T-1| \approx 3H_0 |\vartheta_0'(\vartheta''_0-H_0\vartheta_0')|\left(\frac{\alpha_{L,R}}{\kappa f_{R_0}}\right)^2 < 10^{-15},
\end{equation}
where the above constraint only comes from the $(\alpha k)^2$ contribution (not higher powers in $k$) and the subscript $0$ refers to the present-day values. Note that the leading corrections in Eq.~\eqref{gw: frequency}, which modify the GW speed, are quadratic in $\alpha_{L,R}$, and therefore, do not violate parity symmetry. Assuming that $\vartheta$ evolves on cosmological timescales, such that $\vartheta''\sim H_0\vartheta'$, linear effects in $k\alpha$ (obtained by taking the square root of the above expression) must satisfy
\begin{equation}
\label{eq: ka-2}
    \bigg|\frac{\alpha\vartheta'_0 H_0}{\kappa f_{R_0}}\bigg| < \mathcal{O}(10^{-7}).
\end{equation}
The next-to-leading order correction in $\omega^2_{L,R}$, $(\alpha k)^3$, modifies the dispersion relation, violates parity, and induces velocity birefringence as shown in Fig.~\ref{fig:avb} (right panel). These phase distortions can be constrained with LIGO/Virgo data, as shown in  \cite{Wang:2020cub,Zhao:2022pun}. Using the results from \cite{Zhao:2022pun}, which analyzed the GWTC-3 catalog, one obtains the following bound 
\begin{equation}
\label{eq: ka-3}
    \bigg|\frac{\alpha H_0\vartheta'_0}{f_{R_0}\kappa}\bigg| < \mathcal{O}(10^{-14}).
\end{equation}
A more recent paper \cite{Guo:2025bxz} can also be used to constrain the leading-order velocity birefringence terms $(\alpha k)^3$, which leads to one order of magnitude weaker constraint than the one presented in Eq.~\eqref{eq: ka-3}.

Comparing Eq.~\eqref{eq: ka-2} with \eqref{eq: ka-3}, it is worth noting that, somewhat unexpectedly, higher-order corrections in $\alpha_{L,R}$ appear easier to constrain observationally than lower-order ones, which happens because phase distortions can be measured more precisely than global propagation speeds. We emphasize that both of these constraints apply only to the Palatini formalism, because their effects are absent in the metric formalism.  

Regarding amplitude birefringence, the results do not differ from those of the metric formalism. Using \cite{Lagos:2024boe,Ng:2023jjt} as reference, the friction $\Xi_{L,R}$ in the de-Sitter approximation, given in Eq.~\eqref{eq: de-sitter friction} and assuming that $\vartheta$ evolves on cosmological timescales, along with the low redshift approximation $z \ll 1$, leads to
\begin{equation}
\int\Delta \Xi_{L,R}\mathrm{d}\eta \sim \frac{k \alpha_{L,R} \vartheta_0 H_0^2}{\kappa f_R}\Delta \eta.
\end{equation}
Considering the constraints obtained in \cite{Ng:2023jjt}, we define the following equality and inequality
\begin{equation}
     \frac{k \alpha_{L,R} \vartheta_0 H_0^2}{\kappa f_R}\Delta \eta = \left(\frac{\bar{\alpha}}{\si{Gpc^{-1}}}\right)\left(\frac{f}{100 \si{Hz}}\right)d_c. < \mathcal{O}(1)
\end{equation}
where $\bar{\alpha}$ is the coupling constant used in \cite{Ng:2023jjt} to characterize amplitude birefringence. Noting that $k\sim 2\pi f$ and $\Delta \eta$ is proportional to the comoving distance, the constraint translates into
\begin{equation}
    \bigg|\frac{\alpha \vartheta'_0 H_0}{\kappa f_{R_0}}\bigg| < \mathcal{O}(10^{-5}).
\end{equation}
These results generalize those found in \cite{Sulantay:2022sag,Xiong:2024vsd}, with the inclusion of the $f_{R_0}$ factor.

Recently, the LVK collaboration also reported constraints on GW velocity birefringence \cite{LIGOScientific:2026fcf}. However, a direct translation of these bounds to our model, in either formalism (Palatini or metric), is not possible: the LVK analysis assumes the birefringent signal to be anisotropic and thus the coupling constant is expanded in a spherical-harmonic decomposition on the sky. Our model, by contrast, is formulated on a spatially flat FRW background whose symmetries are both rotations and translations, so the coupling constant is isotropic.

The constraints derived for our gravity model were found under the de-Sitter approximation, which is valid for redshifts $z\lesssim 1$. However, at higher redshifts, the conformal time derivatives of $f(R)$ no longer vanish, and additional contributions to $\omega^2_{L,R}$ and $\Xi_{L,R}$ would need to be included. This may already be important to consider since some GW events can be observed up to $z\sim 1$ \cite{LIGOScientific:2026sit}. Additionally, next-generation detectors, such as LISA \cite{LISA:2024hlh}, Einstein Telescope \cite{ET:2019dnz}, and Cosmic Explorer \cite{Evans:2021gyd}, can observe GW events up to $z\sim 10$. 

We note that complementary constraints on Chern–Simons-type parity violation from multi-messenger and dense-environment probes are typically far more stringent than current GW birefringence bounds, owing to the higher curvature and matter densities sampled in those environments (see related discussion in \cite{Lagos:2024boe}). Whether our model can accommodate a screening mechanism to reconcile such strong local constraints with viable cosmological-scale effects is an open possibility to explore in the future.

\subsection{Birefringence redshift evolution}

Phenomenological models for the evolution of birefringence (amplitude or velocity) as a function of the redshift typically assume that birefringence is proportional to the propagation distance (luminosity or comoving) \cite{Okounkova:2021xjv,Ng:2023jjt,Jenks:2023pmk,Lagos:2024boe}. Current ground-based detectors have already detected GW events up to redshift $z\sim 1$, providing constraints for these phenomenological models. 

Nonetheless, a specific modified model of gravity makes a concrete prediction for this redshift evolution, and it is important to determine whether such predictions are consistent with the phenomenological assumptions adopted in current analyses at $z\sim1$. In this subsection, we illustrate the redshift evolution of birefringence in $f(R)$ gravity during a de-Sitter epoch. Imposing a de-Sitter background is equivalent to requiring a vanishing trace of the metric equation in the Palatini formalism, which gives
\begin{equation}
\label{eq: metric_trace}
    f_R R - 2f = 0, 
\end{equation}
from which it follows that the ratio $f/f_R$ must be approximately equal to the scalar curvature $R$. Note that the explicit form of $f(R)$ has not yet been specified. This is a general result for an arbitrary $f(R)$ function.

We now return to the modified Friedmann equation, Eq.~\eqref{frw equation}. In the de-Sitter limit, the conformal time derivatives on the left-hand side vanish, yielding the simplified expression
\begin{equation}
    \mathcal{H}^2 = \frac{a^2}{6}\left(\frac{\kappa(\rho + 3p)}{f_R} + \frac{f}{f_R}\right).
\end{equation}
Since the accelerated expansion is assumed to be driven by the $f(R)$ contribution rather than the scalar field $\vartheta(\eta)$, it is reasonable to assume that $f(R)$ dominates over the matter-scalar contribution, i.e.,
\begin{equation}
    \kappa (\rho + 3p) \ll f  .
\end{equation}
Under this approximation, the modified Friedmann equation reduces to
\begin{equation}
    \mathcal{H}^2 = \frac{a^2f}{6f_R}.
\end{equation}
Using the metric trace equation, Eq.~\eqref{eq: metric_trace}, the Friedmann equation simplifies greatly to
\begin{equation}
    \mathcal{H}^2 = a^2\left(\frac{R}{12}\right) = a^2H_0^2
\end{equation}
where the scalar curvature $R$ today is expressed in terms of the Hubble constant $H_0$. The above differential equation admits the analytical solution
\begin{equation}
\label{scale factor: de sitter}
    a(\eta) = \frac{-1}{\eta H_0},
\end{equation}
which corresponds to the de-Sitter scale factor in conformal time $\eta$, with $\eta \in (-\infty, 0)$. The relation between the scalar curvature and the Hubble constant then reads
\begin{equation}
    R = 12H_0^2,
\end{equation}
so that, given the current value of the Hubble constant, one can directly determine the corresponding value of $R_0$.

Regarding the scalar field sector, we consider the simplest potential,
\begin{equation}
    V(\vartheta) = V_0 + \frac{1}{2}m^2\vartheta^2,
\end{equation}
where $V_0$ is a constant and $m$ is the mass parameter of the field. This type of potential has been studied in the context of dark energy models; see Refs.~\cite{Wolf:2024eph,Wolf:2025jed}. Note that the scalar field equation depends on the scale factor, which is in principle determined by the choice of $f(R)$. In the present case, we work within a de-Sitter background, where the scale factor can be solved without specifying the explicit form of $f(R)$. Consequently, $f(R)$ does not affect the scalar field dynamics, and the equation of motion reduces to
\begin{equation}
\label{eq:scalar field equation}
    \vartheta'' - \frac{2\vartheta'}{\eta} + 
\frac{\vartheta m^2}{\eta^2 H_0^2} = 0,
\end{equation}
which is the Euler-Cauchy equation, and a discussion of this equation is found in \cite{Mukhanov:2005sc,Weinberg:2008zzc}. This equation admits the analytical solution 
\begin{equation}
    \vartheta(\eta) = \vartheta_1 (-\eta)^{\frac{3}{2} - \frac{3}{2}\sqrt{1 -\frac{4m^2}{9H_0^2}}} + \vartheta_2 (-\eta)^{\frac{3}{2} + \frac{3}{2}\sqrt{1 -\frac{4m^2}{9H_0^2}}},
\end{equation}
where $\vartheta_1$ and $\vartheta_2$ are integration constants. The behavior of the scalar field $\vartheta(\eta)$ depends on the value of $m^2$, and three distinct regimes can be identified. In the regime $9H_0^2 \gg 4m^2$, the scalar field is either a decreasing or increasing function of conformal time, depending on the signs and magnitudes of the integration constants $\vartheta_1$ and $\vartheta_2$. However, regardless of the choice of the integration constants, the scalar field asymptotically tends to a constant as $\eta \to 0$, corresponding to the infinite future in conformal time. In this regime, the exponent of the solution is real; and expanding for $m/H_0 \ll 1$ yields
\begin{equation}
\label{eq: theta real}
    \vartheta(\eta) = \vartheta_1 - \eta^3\vartheta_2 + \mathcal{O}((m/H_0)^2).
\end{equation}
Note that Eq.~\eqref{eq: theta real} confirms that $\vartheta(\eta)$ approaches a constant value as $\eta \to 0$, consistent with the behavior of the full solution. Because the scalar field evolves monotonically and tends to a constant in the infinite future, the overall effect in birefringence would be cumulative over conformal time, as is typically assumed in the literature \cite{Okounkova:2021xjv,Ng:2023jjt,Jenks:2023pmk}. 

For the critical case $9H_0^2 = 4m^2$, the scalar field also evolves monotonically, either decreasing or increasing depending on the sign of the integration constants, and it vanishes as $\eta \to 0$. In this regime, the scalar field takes the form 
\begin{equation}
\label{eq: theta critical}
    \vartheta(\eta) = \vartheta_0(-\eta)^{\frac{3}{2}},
\end{equation}
where $\vartheta_0$ is a linear combination of the original integration constant $\vartheta_1$ and $\vartheta_2$. 

Finally, in the regime $9H_0^2 < 4m^2$, the exponent in the analytical solution becomes complex, leading to a logarithmically oscillatory behavior of the form
\begin{equation}
\label{eq: theta oscillatory}
    \vartheta(\eta) = (-\eta)^{3/2}\left(A\cos\left(\frac{3\mu}{2}\ln(-\eta)\right) + B\sin\left(\frac{3\mu}{2}\ln(-\eta)\right)\right),
\end{equation}
where $\mu = \sqrt{\frac{4m^2}{9H_0^2} - 1}$ and $A$ and $B$ are real constants obtained as linear combinations of $\vartheta_1$ and $\vartheta_2$. In this regime, the scalar field exhibits damped oscillations whose amplitude decays as $\eta \to 0$, as expected for a sufficiently massive field. In this case, the scalar field has an oscillatory behavior as time grows, therefore birefringence also oscillates and does not behave monotonically. 

Next, we shall find an expression for amplitude and velocity birefringence as a function of redshift $z$. By considering that the scalar field behaves monotonically, we will be working with the case $m \ll H_0$, where the scalar field is defined as in Eq.~\eqref{eq: theta real}. 

Having solved the background for the scale factor $a(\eta)$ and the scalar field $\vartheta(\eta)$ we can now completely determine the GW $h_{L,R}(\eta,k)$ defined in Eq.~\eqref{solution gw modified equation}. Note that the expression for the dissipation $\Xi_{L,R}$ is reduced to Eq.~\eqref{eq: de-sitter friction} in a de-Sitter type of space. Replacing the expressions for the scale factor Eq.~\eqref{scale factor: de sitter} and the scalar field Eq.~\eqref{eq: theta real} yields
\begin{equation}
    \Xi_{L,R} \approx -\frac{2}{\eta} + \frac{24k\alpha_{L,R}\eta^3 \vartheta_2 H_0^2}{\kappa f_R},
\end{equation}
where the first term is the prediction coming from GR, and the second term is due to the presence of the dCS $\alpha_{L,R}$ contribution (which we assume to be small, so that its deviation from GR is a small) contribution times the function $f(R)$. As indicated in Eq.~\eqref{solution gw modified equation}, we perform the integration from the time of emission $\eta_e$ to the time of detection today $\eta_d$, leading to 
\begin{equation}
    \int^{\eta_e}_{\eta_d}\Delta\Xi_{L,R}(\hat{\eta},k)\mathrm{d}\hat{\eta} = \left(\frac{6\vartheta_2H_0^2 \alpha_{L,R}k \eta^4}{\kappa f_R}\right)\bigg\vert ^{\eta_e}_{\eta_d},
\end{equation}
Then, transforming from conformal time $\eta$ to redshift $z$ using the relation $\eta \to - \left(\frac{1 + z}{H_0}\right)$, from redshift of emission $z = z_e$ to redshift of detection $z = z_d = 0$, yields the expression
\begin{equation}
    \mu_{A(L,R)} = \frac{6k\alpha_{L,R}\vartheta_2 z_e(4 + 6z_e + 4z_e^2 + z_e^3)}{\kappa f_R H_0^2},
\end{equation}
the term shows dependence on the wave-number $k$ associated with the frequency of the GW $f$, and the redshift $z$. Note that the above expression indicates that amplitude birefringence has up to a cubic dependence on redshift, and it is not proportional to comoving or luminosity distance, which are the standard assumptions in the literature. The coupling constant from dCS induces amplitude birefringence, whereas the introduction of $f(R)$ will enhance or suppress such effect, if $f_R \in ]0,1[$ then it will amplify the phenomena, whereas if $f_R \in [1, \infty[$ it will suppress the contribution. From the above expression we see that 
\begin{equation}
    \mu_{A0 (L,R)} = \frac{6\alpha_{L,R}\vartheta_2 z_e(4 + 6z_e + 4z_e^2 + z_e^3)}{\kappa f_R H_0^2},
\end{equation}
where $\mu_{A0 (L,R)}$ characterizes the strength of amplitude birefringence, as illustrated in Figure \ref{fig:avb} (left plot), where we chose $\mu_{A0} = 10^{-2.3} \rm{s}$ and $\mu_{A0} = 10^{-3.0} \rm{s}$, where the former produces the strongest amplitude birefringence, and the latter produces the smallest deviation from GR. 

By the same token, in de-Sitter space, the angular velocity $\omega^2_{L,R}$ Eq.~\eqref{gw: frequency} is simplified considerably to Eq.\ \eqref{eq: de-sitter omega}. Replacing the scale factor Eq.~\eqref{scale factor: de sitter} and the scalar field Eq.~\eqref{eq: theta real} leads to
\begin{equation}
    \omega^2_{L,R}  \approx k^2 + 162k^2\left(\frac{\alpha^2_{L,R} \eta^6 \vartheta^2_2H_0^4}{\kappa f_R^2}\right) + 999k^3\left(\frac{\alpha^3_{L,R}\eta^{10}\vartheta_2^3H_0^4}{\kappa f_R^3}\right).
\end{equation}
The first term comes from the prediction of GR, whereas the second and third terms are present due to the dCS contribution. The former is parity invariant, whereas the latter violates parity symmetry. Performing the integration from conformal time of emission $\eta_e$ to time of detection today $\eta_t$, as indicated in Eq.~\eqref{solution gw modified equation}, leads to
\begin{align}
    & 16\eta^3H_0^2\left(\frac{\alpha_{L,R}\vartheta_2}{\kappa f_R}\right) \nonumber - \left(\frac{1566\eta^5H_0^4}{5k} + \frac{207k\eta^7H_0^4}{7}\right)\left(\frac{\alpha_{L,R}\vartheta_2}{\kappa f_R}\right)^2 \nonumber \\
    & + \left(1215\eta^9 H_0^6 + \frac{999k^2\eta^{11}H_0^6}{22}\right)\left(\frac{\alpha_{L,R}\vartheta_2}{\kappa f_R}\right)^3\bigg\vert ^{\eta_e}_{\eta_d},
\end{align}
Then transforming from conformal time to redshift,
\begin{align}
&-\frac{16\vartheta_2 P_3(z_e)\alpha_{L,R}}{\kappa f_R H_0} + \frac{9\vartheta_2^2\alpha_{L,R}^2 \left(1218H_0^2 P_5(z_e) + 115k^2 P_7(z_e)\right)}{35k\kappa^2 f_R^2 H_0^3} \nonumber \\
&- \frac{27\vartheta_2^3\alpha_{L,R}^3\left( 990H_0^2 P_9(z_e) + 37k^2 P_{11}(z_e)\right)}{22\kappa^3 f_R^3 H_0^5},
\end{align}
where $P_j(z_e)$ is a polynomial of order $j$ in the redshift variable $z_e$. From the above expression we see that velocity birefringence has a polynomial dependence on redshift, and a contribution up to $k^2$. We define velocity birefringence as a function of its leading contribution in $k$ that breaks parity:
\begin{equation}
    \mu_{V(L,R)} = -\frac{999\vartheta_2^3\alpha_{L,R}^3k^2 P_{11}(z_e)}{22\kappa^3 f_R^3 H_0^5},
\end{equation}
where $\mu_{V0(L,R)}$ is defined as the coefficient
\begin{equation}
    \mu_{V0(L,R)} =-\frac{999\vartheta_2^3\alpha_{L,R}^3P_{11}(z_e)}{22\kappa^3 f_R^3 H_0^5},
\end{equation}
where the function $f(R)$ plays the same role as in amplitude birefringence. In Figure \ref{fig:avb} (right plot), $\mu_{V0} = 10^{-1.35} \si{s^2 \text{rad}}$ and $\mu_{V0} = 10^{-1.45} \si{s^2\text{rad}}$, where the former produces the strongest enhancement of velocity birefringence, while the latter leads to the smallest deviation from GR.

The above characterization of amplitude and velocity birefringence as a function of the redshift was obtained in the Palatini formalism. In the metric formalism, velocity birefringence is absent whereas the amplitude birefringence yields the same result as in the Palatini case. This can be seen by rewriting the modified Friedmann equation (see Appendix \ref{sec:appendix C, metric formalism}) as follows
\begin{equation}
    \mathcal{H}^2 = \frac{a^2}{3f_R}\left(\kappa\rho + \frac{f_R R - f}{2}\right) - \frac{\mathcal{H} f_{RR}R'}{2},
\end{equation}
where we have used $a''/a = \mathcal{H}' + \mathcal{H}^2$ and $R = 6(\mathcal{H}' + \mathcal{H}^2)/a^2$. Imposing a de-Sitter approximation and assuming that the accelerated expansion of the universe is driven by $f(R)$ rather than the scalar field, one obtains
\begin{equation}
    \mathcal{H}^2 = a^2\left(\frac{R}{12}\right),
\end{equation}
which is the same expression as in the Palatini formalism. Consequently, the scale factor takes the same form in both cases. Furthermore, the scalar field obeys the same equation of motion as in the Palatini formalism; as a consequence, the scale factor $a(\eta)$ and the scalar field $\vartheta(\eta)$ have the same form as in both formalisms, and therefore $\Xi_{L,R}$ takes the same form in both formalisms, yielding the same expression for $\mu_{A (L,R)}$.
 
It is worth noting that this equivalence arises only as a result of the de-Sitter approximation, which provides the same background dynamics for the scale factor. Working beyond this approximation, where $R$ is not a constant, the conformal time derivatives of $f(R)$ would no longer vanish, and the dynamics in each formalism would differ. Moreover, the effect of introducing $f(R)$ would not be reduced to a scaling effect.

Finally, this section illustrates the redshift dependence of birefringence (amplitude or velocity), whether in Palatini or metric formalism. In order to obtain explicit expressions, it was necessary to introduce several approximations: first, we work in a de-Sitter universe, defined by a positive constant scalar curvature; second, the potential term is fixed by a quadratic configuration; and third, we assume a small mass. This shows that the redshift evolution of birefringence is highly sensitive to  model assumptions. For this reason, we do not expect to generally have birefringence effects evolving linearly with a cosmological distance (comoving or luminosity) as considered in previous phenomenological models.

\section{Discussion}
\label{sec:discussion}

In this work, we have studied the cosmological propagation of GWs in a spatially-flat FRW background, in conformal time, for an arbitrary function $f(R)$ coupled to a dynamical Chern-Simons term and a scalar field. The equations of motion were derived using both the Palatini and the metric formalisms, each yielding a modified cosmological background. GWs are introduced as linear tensor perturbations of the metric and connection in both formalisms, leading to a modified GW propagation equation. The Palatini formalism predicts amplitude and velocity birefringence, whereas the metric formalism predicts amplitude birefringence only.

To provide further insight into our results, we work within a de-Sitter approximation for our model. In this regime, the conformal-time derivatives of $f(R)$ vanish, simplifying the expression for the GW propagation equation. We find that the effect of the function $f(R)$ is to suppress or enhance the effects of birefringence, depending on whether the value of $f_{R_0}=\partial f/\partial R$ today is larger or smaller than unity.

By considering cosmological models without a cosmological constant, where $f(R)$ drives the accelerated expansion of the universe, we analyze whether the predicted value of $f_{R_0}$ leads to an enhancement or suppression of amplitude and velocity birefringence effects. In order to do this, we analyze three common $f(R)$ functions, namely Hu-Sawicki, Exponential, and  Hyperbolic gravity models, and consider their observational constraints obtained in \cite{Ravi:2024jwl,Plaza:2025gcv}. In the Palatini formalism, all three models predict an enhancement of both amplitude and velocity birefringence, ranging between $1-10\%$ depending on the cosmological data considered. In the metric formalism, only amplitude birefringence is present, and it is also enhanced by up to a factor of 2. 

Having established the predicted enhancement of birefringence for $f(R)$ models of gravity, we now discuss how current GW observations can be used to constrain our model parameters. Using observational constraints on phenomenological models, in which amplitude or velocity birefringence is assumed to be proportional to a cosmological distance, it is possible to translate existing GW bounds into constraints on our model. Amplitude birefringence is  constrained with the GWTC-3 catalog via amplitude waveform distortions.
With regards to velocity modifications, the leading parity-invariant correction $(\alpha k)^2$ is constrained with the multi-messenger event GW170817, whereas the leading parity-violating correction, of order $(\alpha k)^3$, is constrained via phase waveform distortions.  The latter leads to the strongest GW constraints on the CS coupling parameter $\alpha$ so far. Nonetheless, stronger bounds can be obtained from observations in high-density environments, such as the solar system, since CS gravity is not equipped with a screening mechanism.  

Our model of gravity makes theoretical predictions for the birefringence effect as a function of the redshift, and it is important to determine whether such predictions are consistent with current standard assumptions in phenomenological models, where birefringence is assumed to be proportional to a cosmological distance, given that we already detect GW events up to $z\sim 1$. In order to describe the birefringence evolution as a function of the redshift, it is necessary to solve the background equations of the model. To that end, we consider a  quadratic potential for the scalar field and that the accelerated expansion is driven by $f(R)$ rather than the scalar field, in a de-Sitter universe. These assumptions allow us to find exact solutions of the background quantities, namely the scale factor and the scalar field. Under a small scalar mass assumption, we find that amplitude and velocity birefringence have a polynomial dependence on the redshift. The various assumptions made help illustrate that the redshift evolution of birefringence is highly model-dependent and, in general, does not evolve linearly with the comoving or luminosity distance, as typically considered in the literature.

For future work, it would be interesting to study the effects of $f(R)$ gravity in a more general cosmological background, where the conformal-time derivatives of $f(R)$ contribute non-trivially to both the angular velocity and the friction of the GWs in the Palatini formalism, and to compare these results with those of the metric formalism. It would also be worthwhile to study scalar perturbations, particularly their impact on structure formation, and to investigate whether the theory exhibits a screening mechanism capable of evading strong-gravity CS constraints, thereby recovering the predictions of GR in high-density regimes while still predicting observable cosmological effects.

\section{Acknowledgment}
We thank Felipe Sulantay for providing us with material for comparison. We thank Fondecyt Iniciaci\'on 11250105 grant for partially supporting this work. The work of J.P. was funded by the National Agency for Research and Development (ANID)/ Scholarship Program / DOCTORADO NACIONAL/2025 - 21251041.

\appendix

\section{Cosmological solution} 
\label{sec:appendix A, solution}

To solve the perturbed affine connection equations, we express the perturbed connection coefficients in terms of only the background metric and its perturbations. The perturbed connection coefficients are given by:
\begin{widetext}
\begin{align}
    \gamma_{1p}(\eta,k) & = \frac{4\kappa a^2f_R^2\gamma_{4p}a' - 2\kappa a^4f_R^2(\gamma_{2p} + 2h_{p}a') - 2\kappa a^5 h_{p}f_Rf'_R - 2\kappa a^3 f_R^2 (2\gamma'_{3p} + \gamma'_{4p})}{4a^2f_R(\kappa a^2 f_R + \alpha k\vartheta')}   \\
    & + \frac{2\alpha_p k f_R(2\gamma_{3p} - \gamma_{4p})a'\vartheta' + \alpha_p k a(2\gamma_{3p}f'_R - \gamma_{4p}f'_R - 4f_R\gamma'_{3p})\vartheta'}{4a^2f_R(\kappa a^2 f_R + \alpha k\vartheta')}, \\
    \gamma_{2p}(\eta,k) & = \frac{2\kappa a^2f_R\gamma_{4p}a' + \kappa a^4f_Rh'_{p} - \kappa a^3f_R\gamma'_{4p} + 2\alpha_p k\gamma_{4p}a'\vartheta' - \alpha_p k a\gamma_{4p}\vartheta'}{\kappa a^4f_R + \alpha k a^3\vartheta'}, \\
    \gamma_{3p}(\eta,k) & =  \frac{a^3\kappa(-2k\kappa^2f_R^3h_{p} + 4\alpha_p\kappa a^2f_R^2a'h'_{p}\vartheta' + 2\alpha_p\kappa a^3f_R\vartheta'(-k^2h_{p}f_R + h'_{p}f'_R) + 2\alpha_p^2 k f_R a'\vartheta'^2h'_{p})}{4k(\kappa a^2f_R + \alpha k\vartheta')^3} \nonumber\\
    & + \frac{\alpha_p^2kah'_{p}f'_R\vartheta'^2}{4k(\kappa a^2f_R + \alpha k\vartheta')^3}, \\
    \gamma_{4p}(\eta,k) & = \frac{a^3\kappa(2k\kappa a^3f_R^2h_{p} - 2\alpha_p f_Ra'h'_{p}\vartheta' + \alpha_p a(2k^2f_Rh_{p} - h'_{p}f'_R)\vartheta'}{2k(\kappa a^2f_R + \alpha k\vartheta')^2},
\end{align}
\end{widetext}
where $\gamma_{1p}(\eta,k)$ and $\gamma_{2p}(\eta,k)$ are written in terms of $\gamma_{3p}(\eta,k)$ and $\gamma_{4p}(\eta,k)$.

\section{Cosmological series expansion} 
\label{sec:appendix B, series expansion}

The coefficients $A_{mn}$, $B_{mn}$, $C_{mn}$, and $D_{mn}$ are expressed solely as  functions of background quantities. The $A_{mn}$ coefficients are
\begin{align*}
A_{00} & = 4\kappa^4 a^{11}f^6_R, \\
A_{11} & = 12\kappa^3 a^{9}f^5_R\vartheta', \\
A_{20} & = -12 \kappa ^2 a^6 f^3_R a' \vartheta '^2 f'_R-12 \kappa ^2 a^5 f_R^4 a'^2 \vartheta '^2 - 3 \kappa ^2 a^7 f^2_R \vartheta '^2 f'^2_R, \\
A_{22} & = 12 \kappa^2 a^7 f^4_R\vartheta'^2, \\
A_{31} & = -20 \kappa a^4 f_R^2 a' \vartheta '^3 f'_R - 20 \kappa  a^3 f_R^3 a'^2 \vartheta '^3-5 \kappa  a^5 f_R \vartheta '^3 f'^2_R, \\
A_{33} & = 4\kappa a^5 f^3_R \vartheta'^3,\\
A_{42} & = -8 a^2 f_R a' \vartheta '^4 f'_R -8 a f_R^2 a'^2 \vartheta '^4-2 a^3 \vartheta '^4 f'^2_R.
\end{align*}

The $B_{mm}$ coefficients are
\begin{align*}
B_{00} & = 8\kappa^4 a^{10}f_R^6a' + 4\kappa^4 a^{11}f_R^5f'_R,\\
B_{11} & = 40 \kappa ^3 a^8 f_R^5 a' \vartheta '+20 \kappa ^3 a^9 f_R^4 \vartheta ' f'_R -8 \kappa ^3 a^9 f_R^5 \vartheta '', \\
B_{20} & = -12 \kappa ^2 a^6 f_R^3 a'' \vartheta '^2 f'_R-12 \kappa ^2 a^6 f_R^3 a' \vartheta '^2 f_R'' \\
&+30 \kappa ^2 a^6 f_R^2 a' \vartheta '^2 f'^2_R+48 \kappa ^2 a^5 f_R^3 a'^2 \vartheta '^2 f'_R \\
&-24 \kappa ^2 a^6 f_R^3 a' \vartheta ' \vartheta '' f'_R+48 \kappa ^2 a^4 f_R^4 a'^3 \vartheta '^2 \\
&-24 \kappa ^2 a^5 f_R^4 a'^2 \vartheta ' \vartheta ''-24 \kappa ^2 a^5 f_R^4 a' a'' \vartheta '^2+9 \kappa ^2 a^7 f_R\vartheta '^2 f'^3_R \\
&-6 \kappa ^2 a^7 f_R^2 \vartheta ' \vartheta '' f'^2_R-6 \kappa ^2 a^7 f_R^2 \vartheta '^2 f'_R f''_R,\\
B_{22} & = 56 \kappa ^2 a^6 f_R^4 a' \vartheta '^2+28 \kappa ^2 a^7 f_R^3 \vartheta '^2 f'_R-16 \kappa ^2 a^7 f_R^4 \vartheta ' \vartheta '', \\
B_{31} & = -20 \kappa  a^4 f_R^2 a'' \vartheta '^3 f'_R-20 \kappa  a^4 f_R^2 a' \vartheta '^3 f''_R \\
&-10 \kappa  a^4 f_R a' \vartheta '^3 f'^2_R-40 \kappa  a^3 f_R^2 a'^2 \vartheta '^3 f'_R-40 \kappa  a^3 f_R^3 a' a'' \vartheta '^3 \\
&+5 \kappa  a^5 \vartheta '^3 f'^3_R-10 \kappa  a^5 f_R \vartheta '^3 f'_R f''_R,\\
B_{33} & = 24 \kappa  a^4 f_R^3 a' \vartheta '^3+12 \kappa  a^5 f_R^2 \vartheta '^3 f'_R-8 \kappa  a^5 f_R^3 \vartheta '^2 \vartheta '',\\
B_{42} & = -8 a^2 f_R a'' \vartheta '^4 f'_R-8 a^2 f_R a' \vartheta '^4 f''_R-16 a^2 a' \vartheta '^4 f'^2_R \\
&-40 a f_R a'^2 \vartheta '^4 f'_R+8 a^2 f_R a' \vartheta '^3 \vartheta '' f'_R-16 f_R^2 a'^3 \vartheta '^4 \\
& +8 a f_R^2 a'^2 \vartheta '^3 \vartheta ''-16 a f_R^2 a' a'' \vartheta '^4+2 a^3 \vartheta '^3 \vartheta '' f'^2_R-4 a^3 \vartheta '^4 f'_R f''_R.
\end{align*}

The $C_{mn}$ coefficients are
\begin{align*}
C_{02} & = 4\kappa^4 a^{10} f^5_R, \\
C_{13} & = 8\kappa^3 a^8 f^4_R \vartheta',\\
C_{22} & = -12 \kappa ^2 a^5 f_R^3 a'' \vartheta'^2+24 \kappa ^2 a^5 f_R^2 a' \vartheta '^2 f'_R + 36 \kappa ^2 a^4 f_R^3 a'^2 \vartheta '^2 \\
&-24 \kappa ^2 a^5 f_R^3 a' \vartheta ' \vartheta ''-6 \kappa ^2 a^6 f_R^2 \vartheta '^2 f''_R+12 \kappa ^2 a^6 f_R \vartheta '^2 f'^2_R \\
&-12 \kappa ^2 a^6 f_R^2 \vartheta ' \vartheta '' f'_R, \\
C_{24} & =  4\kappa^2 a^6 f^3_R\vartheta'^2,\\
C_{33} & = -20 \kappa  a^3 f_R^2 a'' \vartheta '^3-16 \kappa  a^3 f_R a'\vartheta '^3 f'_R+4 \kappa  a^2 f_R^2 a'^2 \vartheta '^3 \\
&-12 \kappa  a^3 f_R^2 a'\vartheta '^2 \vartheta ''-10 \kappa  a^4 f_R \vartheta '^3 f''_R+6 \kappa  a^4 \vartheta '^3 f'^2_R, \\
&-6 \kappa  a^4 f_R \vartheta '^2 \vartheta '' f'_R,\\
C_{44} & = -8 a f_R a'' \vartheta '^4-16 a a' \vartheta '^4 f'_R- 8f_R a'^2 \vartheta '^4-4 a^2 \vartheta '^4 f''_R.
\end{align*}

The $D_{mn}$ coefficients are
\begin{align*}
D_{00} & = 4\kappa^4 a^{10}f^5_R, \\
D_{11} & = 8\kappa^3 a^{8}f^4_R\vartheta', \\
D_{20} & = -12 \kappa ^2 a^4 f^3_R a'^2 \vartheta '^2 - 12 \kappa ^2 a^5 f_R^2 a' f'_R \vartheta '^2 - 3 \kappa ^2 a^6 f_R \vartheta '^2 f'^2_R, \\
D_{22} & = 4 \kappa^2 a^6 f^3_R\vartheta'^2, \\
D_{31} & = -8 \kappa  a^2 f^2_R a'^2 \vartheta '^3 - 8 \kappa a^3 f_R a' f'_R \vartheta '^3 - 2 \kappa  a^4  \vartheta '^3 f'^2_R. 
\end{align*}

\section{Metric formalism} 
\label{sec:appendix C, metric formalism}

In this appendix, we briefly show how to obtain the gravitational wave equation for $f(R)$ gravity coupled to a dCS term using the metric formalism, where the manifold is endowed with a metric structure and the connection is the Levi-Civita connection. A variation of the action, Eq.\ \eqref{eq: action}, with respect to the inverse metric tensor yields
\begin{equation}
    f_R R_{\mu\nu} - \frac{1}{2}fg_{\mu\nu} + g_{\mu\nu}\square f_R - \nabla_\mu\nabla_\nu f_R + \frac{\alpha}{\kappa}C_{\mu\nu} = T^{\text{mat}}_{\mu\nu} + T^{\vartheta}_{\mu\nu},
\end{equation}
where the matter field is modeled as a perfect fluid, Eq.\ \eqref{FRWfluid}, and the scalar field stress-energy tensor is defined in Eq.\ \eqref{eq: scalar field tensor}. The $C_{\mu\nu}$ tensor is defined as 
\begin{equation}
    C^{\mu\nu} = \frac{1}{\sqrt{-g}}\left(\left(\nabla_\rho\vartheta\right)\epsilon^{\rho\sigma\lambda(\mu}\nabla_\lambda R^{\nu)}{}_{\sigma} + \left(\nabla_\rho\nabla_\sigma \vartheta\right)^*R^{\sigma (\mu\nu)\rho}\right).
\end{equation}

Substituting the metric tensor, Eq.~\eqref{conformal metric}, the scalar field, Eq.~\eqref{cosmo scalar}, and the energy-momentum tensor, Eq.~\eqref{FRWfluid}, into the field equations leads to
\begin{align}
    \kappa a^2\rho - \frac{1}{2}a^2f - 3f_R\mathcal{H}^2 + 3 f_R\frac{a''}{a} - 3f_R'\mathcal{H} & = 0, \\
    -\kappa a^2 p -\frac{1}{2}a^2f + f_R\mathcal{H}^2 + f_R\frac{a''}{a} - f'_R\mathcal{H} - f''_R& = 0.
\end{align}
The scalar field equation and the continuity equation for the matter contribution remain the same as those in the Palatini formalism. 

Introducing the tensor perturbations of the metric tensor as in Eq.~\eqref{line element}, and changing the polarization basis from $(h_+,h_\times)$ to $(h_L,h_R)$ in Fourier space, one obtains the modified GW equation in the metric formalism:
\begin{equation}
\label{eq: gw metric}
    h''_{L,R} + \left(\frac{2\mathcal{H} + \frac{f'_R}{f_R} + \frac{2\alpha_{L,R} k\vartheta''}{a^2 \kappa f_R}}{1 + \frac{2\alpha_{L,R} k\vartheta'}{a^2\kappa f_R}}\right)h'_{L,R} + k^2h_{L,R} = 0,
\end{equation}
where the friction term of the GW equation is given by
\begin{equation}
    \Xi_{L,R} = \frac{2\mathcal{H} + \frac{f'_R}{f_R} + \frac{2\alpha_{L,R} k\vartheta''}{a^2 \kappa f_R}}{1 + \frac{2\alpha_{L,R} k\vartheta'}{a^2\kappa f_R}},
\end{equation}
which is in agreement with \cite{Odintsov:2022hxu}. Treating $\alpha_{L,R}$ as a small parameter, one finds that $\Xi_{L,R}$ takes the same form as the expression obtained in the Palatini formalism.  It can be seen from Eq.\ \eqref{eq: gw metric} that the GW amplitude is polarization-dependent, yielding amplitude birefringence; however, it is clear from the same equation that velocity birefringence is absent.

\bibliographystyle{apsrev4-1}
\bibliography{refs.bib}

\end{document}